\input harvmac
\let\includefigures=\iftrue
\let\useblackboard==\iftrue
\newfam\black

\includefigures
\message{If you do not have epsf.tex (to include figures),}
\message{change the option at the top of the tex file.}
\def\figin{\epsfcheck\figin}\def\figins{\epsfcheck\figins}
\def\epsfcheck{\ifx\epsfbox\UnDeFiNeD
\message{(NO epsf.tex, FIGURES WILL BE IGNORED)}
\gdef\figin##1{\vskip2in}\gdef\figins##1{\hskip.5in}
\else\message{(FIGURES WILL BE INCLUDED)}%
\gdef\figin##1{##1}\gdef\figins##1{##1}\fi}
\def\DefWarn#1{}
\def\figinsert{\goodbreak\midinsert}
\def\ifig#1#2#3{\DefWarn#1\xdef#1{fig.~\the\figno}
\writedef{#1\leftbracket fig.\noexpand~\the\figno}%
\figinsert\figin{\centerline{#3}}\medskip\centerline{\vbox{
\baselineskip12pt\advance\hsize by -1truein
\noindent\footnotefont{\bf Fig.~\the\figno:} #2}}
\endinsert\global\advance\figno by1}
\else
\def\ifig#1#2#3{\xdef#1{fig.~\the\figno}
\writedef{#1\leftbracket fig.\noexpand~\the\figno}%
\global\advance\figno by1} \fi
\def\id{{1 \kern-.28em {\rm l}}}

\def\K3{{\bf K3}}
\def\journal#1&#2(#3){\unskip, \sl #1\ \bf #2 \rm(19#3) }
\def\andjournal#1&#2(#3){\sl #1~\bf #2 \rm (19#3) }

\def\bar{\overline}
\def\hat{\widehat}
\def\ie{{\it i.e.}}
\def\eg{{\it e.g.}}

\def\frac#1#2{{#1\over#2}}

\def\half{\frac12}

\def\inbar{\,\vrule height1.5ex width.4pt depth0pt}
\def\IC{\relax\hbox{$\inbar\kern-.3em{\rm C}$}}
\def\IR{\relax{\rm I\kern-.18em R}}
\def\IZ{\relax{\rm I\kern-.18em Z}}

%
%

%
\catcode`\@=11
\def\slash#1{\mathord{\mathpalette\c@ncel{#1}}}
\overfullrule=0pt

\def\DD{{\cal D}}

\def\LL{{\cal L}}
\def\MM{{\cal M}}
\def\NN{{\cal N}}
\def\OO{{\cal O}}

\def\underrel#1\over#2{\mathrel{\mathop{\kern\z@#1}\limits_{#2}}}

\catcode`\@=12


%

\def\det{{\rm det}}

\def\det{{\rm det}}
\def\exp{{\rm exp}}


\def\ie{{\it i.e.}}
\def\eg{{\it e.g.}}


\lref\CompereJK{
  G.~Compère,
  ``The Kerr/CFT correspondence and its extensions,''
Living Rev.\ Rel.\  {\bf 15}, 11 (2012), [Living Rev.\ Rel.\  {\bf 20}, no. 1, 1 (2017)].
[arXiv:1203.3561 [hep-th]].
}

\lref\LindbladNigel{
G.~Lindblad and B.~Nagel,
``Continuous bases for unitary irreducible representations of SU(1,1),''
Annales de l'I.H.P. Physique théorique,  Volume 13 (1970) no. 1,  p. 27-56.
}

\lref\GiveonCC{
  A.~Giveon and E.~Witten,
  ``Mirror symmetry as a gauge symmetry,''
Phys.\ Lett.\ B {\bf 332}, 44 (1994).
[hep-th/9404184].
}

\lref\GiveonKU{
  A.~Giveon and A.~Pakman,
  ``More on superstrings in AdS(3) x N,''
JHEP {\bf 0303}, 056 (2003).
[hep-th/0302217].
}

\lref\KutasovJP{
  D.~Kutasov and D.~A.~Sahakyan,
  ``Comments on the thermodynamics of little string theory,''
JHEP {\bf 0102}, 021 (2001).
[hep-th/0012258].
}

\lref\ForsteWP{
  S.~Forste,
  ``A Truly marginal deformation of SL(2, R) in a null direction,''
Phys.\ Lett.\ B {\bf 338}, 36 (1994).
[hep-th/9407198].
}

\lref\GiveonJG{
  A.~Giveon and M.~Rocek,
  ``Supersymmetric string vacua on AdS(3) x N,''
JHEP {\bf 9904}, 019 (1999).
[hep-th/9904024].
}

\lref\BerensteinGJ{
  D.~Berenstein and R.~G.~Leigh,
  ``Space-time supersymmetry in AdS(3) backgrounds,''
Phys.\ Lett.\ B {\bf 458}, 297 (1999).
[hep-th/9904040].
}

\lref\GiveonFGR{
  A.~Giveon,
  ``Comments on $T\bar T$, $J\bar{T}$ and String Theory,''
[arXiv:1903.06883 [hep-th]].
}

\lref\KutasovZH{
  D.~Kutasov, F.~Larsen and R.~G.~Leigh,
  ``String theory in magnetic monopole backgrounds,''
Nucl.\ Phys.\ B {\bf 550}, 183 (1999).
[hep-th/9812027].
}

\lref\ElShowkCM{
  S.~El-Showk and M.~Guica,
  ``Kerr/CFT, dipole theories and nonrelativistic CFTs,''
JHEP {\bf 1212}, 009 (2012).
[arXiv:1108.6091 [hep-th]].
}

\lref\CompereJK{
  G.~Compère,
  ``The Kerr/CFT correspondence and its extensions,''
Living Rev.\ Rel.\  {\bf 15}, 11 (2012), [Living Rev.\ Rel.\  {\bf 20}, no. 1, 1 (2017)].
[arXiv:1203.3561 [hep-th]].
}

\lref\deBoerGYT{
  J.~de Boer, H.~Ooguri, H.~Robins and J.~Tannenhauser,
  ``String theory on AdS(3),''
JHEP {\bf 9812}, 026 (1998).
[hep-th/9812046].
}

\lref\GiveonKU{
  A.~Giveon and A.~Pakman,
  ``More on superstrings in AdS(3) x N,''
JHEP {\bf 0303}, 056 (2003).
[hep-th/0302217].
}

\lref\dattajiang{S. Chakraborty, S. Datta, A. Giveon, Y. Jiang and D. Kutasov, in progress.}

\lref\CooperFFA{
  P.~Cooper, S.~Dubovsky and A.~Mohsen,
  ``Ultraviolet complete Lorentz-invariant theory with superluminal signal propagation,''
Phys.\ Rev.\ D {\bf 89}, no. 8, 084044 (2014).
[arXiv:1312.2021 [hep-th]].
}

\lref\BzowskiPCY{
  A.~Bzowski and M.~Guica,
  ``The holographic interpretation of $J \bar T$-deformed CFTs,''
[arXiv:1803.09753 [hep-th]].
}

\lref\DijkgraafVV{
  R.~Dijkgraaf, E.~P.~Verlinde and H.~L.~Verlinde,
  ``Matrix string theory,''
Nucl.\ Phys.\ B {\bf 500}, 43 (1997).
[hep-th/9703030].
}

\lref\GiveonFU{
  A.~Giveon, M.~Porrati and E.~Rabinovici,
  ``Target space duality in string theory,''
Phys.\ Rept.\  {\bf 244}, 77 (1994).
[hep-th/9401139].
}

\lref\ParsonsSI{
  J.~Parsons and S.~F.~Ross,
  ``Strings in extremal BTZ black holes,''
JHEP {\bf 0904}, 134 (2009).
[arXiv:0901.3044 [hep-th]].
}

\lref\BarsSR{
  I.~Bars and K.~Sfetsos,
  ``Conformally exact metric and dilaton in string theory on curved space-time,''
Phys.\ Rev.\ D {\bf 46}, 4510 (1992).
[hep-th/9206006].
}

\lref\DetournayFZ{
  S.~Detournay, D.~Orlando, P.~M.~Petropoulos and P.~Spindel,
  ``Three-dimensional black holes from deformed anti-de Sitter,''
JHEP {\bf 0507}, 072 (2005).
[hep-th/0504231].
}

\lref\AzeyanagiZD{
  T.~Azeyanagi, D.~M.~Hofman, W.~Song and A.~Strominger,
  ``The Spectrum of Strings on Warped $AdS_3 \times S^3$,''
JHEP {\bf 1304}, 078 (2013).
[arXiv:1207.5050 [hep-th]].
}

\lref\SmirnovLQW{
  F.~A.~Smirnov and A.~B.~Zamolodchikov,
  ``On space of integrable quantum field theories,''
Nucl.\ Phys.\ B {\bf 915}, 363 (2017).
[arXiv:1608.05499 [hep-th]].
}

\lref\CavagliaODA{
  A.~Cavaglià, S.~Negro, I.~M.~Szécsényi and R.~Tateo,
  ``$T \bar{T}$-deformed 2D Quantum Field Theories,''
JHEP {\bf 1610}, 112 (2016).
[arXiv:1608.05534 [hep-th]].
}

\lref\LeFlochRUT{
  B.~Le Floch and M.~Mezei,
  ``Solving a family of $T\bar{T}$-like theories,''
[arXiv:1903.07606 [hep-th]].
}

\lref\McGoughLOL{
  L.~McGough, M.~Mezei and H.~Verlinde,
  ``Moving the CFT into the bulk with $ T\overline{T} $,''
JHEP {\bf 1804}, 010 (2018).
[arXiv:1611.03470 [hep-th]].
}

\lref\GiveonNIE{
  A.~Giveon, N.~Itzhaki and D.~Kutasov,
  ``$ T\bar{T} $ and LST,''
JHEP {\bf 1707}, 122 (2017).
[arXiv:1701.05576 [hep-th]].
}

\lref\GiveonMYJ{
  A.~Giveon, N.~Itzhaki and D.~Kutasov,
  ``A solvable irrelevant deformation of AdS$_{3}$/CFT$_{2}$,''
JHEP {\bf 1712}, 155 (2017).
[arXiv:1707.05800 [hep-th]].
}

\lref\ChakrabortyVJA{
  S.~Chakraborty, A.~Giveon and D.~Kutasov,
  ``$ J\overline{T} $ deformed CFT$_{2}$ and string theory,''
JHEP {\bf 1810}, 057 (2018).
[arXiv:1806.09667 [hep-th]].
}

\lref\ApoloQPQ{
  L.~Apolo and W.~Song,
  ``Strings on warped AdS$_{3}$ via $ \mathrm{T}\bar{\mathrm{J}} $ deformations,''
JHEP {\bf 1810}, 165 (2018).
[arXiv:1806.10127 [hep-th]].
}

\lref\KutasovXB{
  D.~Kutasov,
  ``Geometry on the Space of Conformal Field Theories and Contact Terms,''
Phys.\ Lett.\ B {\bf 220}, 153 (1989)..
}

\lref\PolchinskiRQ{
  J.~Polchinski,
  ``String theory. Vol. 1: An introduction to the bosonic string,''
}

\lref\GiveonMYJ{
  A.~Giveon, N.~Itzhaki and D.~Kutasov,
  ``A solvable irrelevant deformation of AdS$_{3}$/CFT$_{2}$,''
JHEP {\bf 1712}, 155 (2017).
[arXiv:1707.05800 [hep-th]].
}

\lref\MaldacenaHW{
  J.~M.~Maldacena and H.~Ooguri,
  ``Strings in AdS(3) and SL(2,R) WZW model 1: The Spectrum,''
J.\ Math.\ Phys.\  {\bf 42}, 2929 (2001).
[hep-th/0001053].
}

\lref\ShyamZNQ{
  V.~Shyam,
  ``Background independent holographic dual to $T\bar{T}$ deformed CFT with large central charge in 2 dimensions,''
JHEP {\bf 1710}, 108 (2017).
[arXiv:1707.08118 [hep-th]].
}

\lref\AsratTZD{
  M.~Asrat, A.~Giveon, N.~Itzhaki and D.~Kutasov,
  ``Holography Beyond AdS,''
[arXiv:1711.02690 [hep-th]].
}

\lref\GiribetIMM{
  G.~Giribet,
  ``$T\bar{T}$-deformations, AdS/CFT and correlation functions,''
JHEP {\bf 1802}, 114 (2018).
[arXiv:1711.02716 [hep-th]].
}

\lref\KrausXRN{
  P.~Kraus, J.~Liu and D.~Marolf,
  ``Cutoff AdS$_3$ versus the $T\bar{T}$ deformation,''
[arXiv:1801.02714 [hep-th]].
}

\lref\CardySDV{
  J.~Cardy,
  ``The $T\overline T$ deformation of quantum field theory as a stochastic process,''
[arXiv:1801.06895 [hep-th]].
}

\lref\CottrellSKZ{
  W.~Cottrell and A.~Hashimoto,
  ``Comments on $T \bar T$ double trace deformations and boundary conditions,''
[arXiv:1801.09708 [hep-th]].
}

\lref\AharonyVUX{
  O.~Aharony and T.~Vaknin,
  ``The TT* deformation at large central charge,''
[arXiv:1803.00100 [hep-th]].
}

\lref\DubovskyDLK{
  S.~Dubovsky,
  ``A Simple Worldsheet Black Hole,''
[arXiv:1803.00577 [hep-th]].
}

\lref\BonelliKIK{
  G.~Bonelli, N.~Doroud and M.~Zhu,
  ``$T\bar T$-deformations in closed form,''
[arXiv:1804.10967 [hep-th]].
}

\lref\BaggioGCT{
  M.~Baggio and A.~Sfondrini,
  ``Strings on NS-NS Backgrounds as Integrable Deformations,''
[arXiv:1804.01998 [hep-th]].
}

\lref\ChakrabortyKPR{
  S.~Chakraborty, A.~Giveon, N.~Itzhaki and D.~Kutasov,
  ``Entanglement Beyond $\rm AdS$,''
[arXiv:1805.06286 [hep-th]].
}

\lref\GuicaLIA{
  M.~Guica,
  ``An integrable Lorentz-breaking deformation of two-dimensional CFTs,''
[arXiv:1710.08415 [hep-th]].
}

\lref\BzowskiPCY{
  A.~Bzowski and M.~Guica,
  ``The holographic interpretation of $J \bar T$-deformed CFTs,''
[arXiv:1803.09753 [hep-th]].
}

\lref\StromingerCZ{
  A.~Strominger,
  ``Massless black holes and conifolds in string theory,''
Nucl.\ Phys.\ B {\bf 451}, 96 (1995).
[hep-th/9504090].
}

\lref\IRownNW{
  J.~D.~Brown and M.~Henneaux,
  ``Central Charges in the Canonical Realization of Asymptotic Symmetries: An Example from Three-Dimensional Gravity,''
Commun.\ Math.\ Phys.\  {\bf 104}, 207 (1986)..
}

\lref\GiveonCGS{
  A.~Giveon and D.~Kutasov,
  ``Supersymmetric Renyi entropy in CFT$_{2}$ and AdS$_{3}$,''
JHEP {\bf 1601}, 042 (2016).
[arXiv:1510.08872 [hep-th]].
}

\lref\KutasovXQ{
  D.~Kutasov and A.~Schwimmer,
  ``Universality in two-dimensional gauge theory,''
Nucl.\ Phys.\ B {\bf 442}, 447 (1995).
[hep-th/9501024].
}

\lref\GiveonNS{
  A.~Giveon, D.~Kutasov and N.~Seiberg,
  ``Comments on string theory on AdS(3),''
Adv.\ Theor.\ Math.\ Phys.\  {\bf 2}, 733 (1998).
[hep-th/9806194].
}

\lref\KutasovXU{
  D.~Kutasov and N.~Seiberg,
  ``More comments on string theory on AdS(3),''
JHEP {\bf 9904}, 008 (1999).
[hep-th/9903219].
}

\lref\SeibergXZ{
  N.~Seiberg and E.~Witten,
  ``The D1 / D5 system and singular CFT,''
JHEP {\bf 9904}, 017 (1999).
[hep-th/9903224].
}

\lref\GiveonUP{
  A.~Giveon and D.~Kutasov,
  ``Notes on AdS(3),''
Nucl.\ Phys.\ B {\bf 621}, 303 (2002).
[hep-th/0106004].
}

\lref\ArgurioTB{
  R.~Argurio, A.~Giveon and A.~Shomer,
  ``Superstrings on AdS(3) and symmetric products,''
JHEP {\bf 0012}, 003 (2000).
[hep-th/0009242].
}

\lref\GiveonMI{
  A.~Giveon, D.~Kutasov, E.~Rabinovici and A.~Sever,
  ``Phases of quantum gravity in AdS(3) and linear dilaton backgrounds,''
Nucl.\ Phys.\ B {\bf 719}, 3 (2005).
[hep-th/0503121].
}

\lref\MotlTH{
  L.~Motl,
  ``Proposals on nonperturbative superstring interactions,''
[hep-th/9701025].
}

\lref\DijkgraafVV{
  R.~Dijkgraaf, E.~P.~Verlinde and H.~L.~Verlinde,
  ``Matrix string theory,''
Nucl.\ Phys.\ B {\bf 500}, 43 (1997).
[hep-th/9703030].
}

\lref\GiveonPX{
  A.~Giveon and D.~Kutasov,
  ``Little string theory in a double scaling limit,''
JHEP {\bf 9910}, 034 (1999).
[hep-th/9909110].
}

\lref\GiveonCGS{
  A.~Giveon and D.~Kutasov,
  ``Supersymmetric Renyi entropy in CFT$_{2}$ and AdS$_{3}$,''
JHEP {\bf 1601}, 042 (2016).
[arXiv:1510.08872 [hep-th]].
}

\lref\KutasovXB{
  D.~Kutasov,
  ``Geometry on the Space of Conformal Field Theories and Contact Terms,''
Phys.\ Lett.\ B {\bf 220}, 153 (1989).
}

\lref\GiveonZM{
  A.~Giveon, D.~Kutasov and O.~Pelc,
  ``Holography for noncritical superstrings,''
JHEP {\bf 9910}, 035 (1999).
[hep-th/9907178].
}

\lref\ZamolodchikovBD{
  A.~B.~Zamolodchikov and V.~A.~Fateev,
  ``Operator Algebra and Correlation Functions in the Two-Dimensional Wess-Zumino SU(2) x SU(2) Chiral Model,''
Sov.\ J.\ Nucl.\ Phys.\  {\bf 43}, 657 (1986), [Yad.\ Fiz.\  {\bf 43}, 1031 (1986)].
}

\lref\TeschnerFT{
  J.~Teschner,
  ``On structure constants and fusion rules in the SL(2,C) / SU(2) WZNW model,''
Nucl.\ Phys.\ B {\bf 546}, 390 (1999). [hep-th/9712256]; A. B. Zamolodchikov and Al. B.
Zamolodchikov, unpublished.
}

\lref\MotlTH{
  L.~Motl,
  ``Proposals on nonperturbative superstring interactions,''
[hep-th/9701025].
}

\lref\KlemmDF{
  A.~Klemm and M.~G.~Schmidt,
  ``Orbifolds by Cyclic Permutations of Tensor Product Conformal Field Theories,''
Phys.\ Lett.\ B {\bf 245}, 53 (1990).
}

\lref\FuchsVU{
  J.~Fuchs, A.~Klemm and M.~G.~Schmidt,
  ``Orbifolds by cyclic permutations in Gepner type superstrings and in the corresponding Calabi-Yau manifolds,''
Annals Phys.\  {\bf 214}, 221 (1992).
}

\lref\WakimotoGF{
  M.~Wakimoto,
  ``Fock representations of the affine lie algebra A1(1),''
Commun.\ Math.\ Phys.\  {\bf 104}, 605 (1986).
}

\lref\DubovskyCNJ{
  S.~Dubovsky, V.~Gorbenko and M.~Mirbabayi,
  ``Asymptotic fragility, near AdS$_{2}$ holography and $ T\overline{T} $,''
JHEP {\bf 1709}, 136 (2017).
[arXiv:1706.06604 [hep-th]].
}

\lref\BernardIY{
  D.~Bernard and G.~Felder,
  ``Fock Representations and BRST Cohomology in SL(2) Current Algebra,''
Commun.\ Math.\ Phys.\  {\bf 127}, 145 (1990).
}

\lref\DubovskyWK{
  S.~Dubovsky, R.~Flauger and V.~Gorbenko,
  ``Solving the Simplest Theory of Quantum Gravity,''
JHEP {\bf 1209}, 133 (2012).
[arXiv:1205.6805 [hep-th]].
}

\lref\BershadskyIN{
  M.~Bershadsky and D.~Kutasov,
  ``Comment on gauged WZW theory,''
Phys.\ Lett.\ B {\bf 266}, 345 (1991).
}

\lref\ElitzurRT{
  S.~Elitzur, A.~Giveon, D.~Kutasov and E.~Rabinovici,
  ``From big bang to big crunch and beyond,''
JHEP {\bf 0206}, 017 (2002).
[hep-th/0204189].
}

\lref\CrapsII{
  B.~Craps, D.~Kutasov and G.~Rajesh,
  ``String propagation in the presence of cosmological singularities,''
JHEP {\bf 0206}, 053 (2002).
[hep-th/0205101].
}

\lref\ZamolodchikovCE{
  A.~B.~Zamolodchikov,
  ``Expectation value of composite field T anti-T in two-dimensional quantum field theory,''
[hep-th/0401146].
}

\lref\CaselleDRA{
  M.~Caselle, D.~Fioravanti, F.~Gliozzi and R.~Tateo,
  ``Quantisation of the effective string with TBA,''
JHEP {\bf 1307}, 071 (2013).
[arXiv:1305.1278 [hep-th]].
}

\lref\DattaTHY{
  S.~Datta and Y.~Jiang,
  ``$T\bar{T}$ deformed partition functions,''
[arXiv:1806.07426 [hep-th]].
}

\lref\IsraelVV{
  D.~Israel, C.~Kounnas, D.~Orlando and P.~M.~Petropoulos,
  ``Electric/magnetic deformations of S**3 and AdS(3), and geometric cosets,''
Fortsch.\ Phys.\  {\bf 53}, 73 (2005).
[hep-th/0405213].
}

\lref\DetournayRH{
  S.~Detournay, D.~Israel, J.~M.~Lapan and M.~Romo,
  ``String Theory on Warped $AdS_{3}$ and Virasoro Resonances,''
JHEP {\bf 1101}, 030 (2011).
[arXiv:1007.2781 [hep-th]].
}

\lref\DubovskyIRA{
  S.~Dubovsky, V.~Gorbenko and M.~Mirbabayi,
  ``Natural Tuning: Towards A Proof of Concept,''
JHEP {\bf 1309}, 045 (2013).
[arXiv:1305.6939 [hep-th]].
}

\lref\StromingerEQ{
  A.~Strominger,
  ``Black hole entropy from near horizon microstates,''
JHEP {\bf 9802}, 009 (1998).
[hep-th/9712251].
}

\lref\GiveonPR{
  A.~Giveon and D.~Kutasov,
  ``Fundamental strings and black holes,''
JHEP {\bf 0701}, 071 (2007).
[hep-th/0611062].
}

\lref\DubovskyCNJ{
  S.~Dubovsky, V.~Gorbenko and M.~Mirbabayi,
  ``Asymptotic fragility, near AdS$_{2}$ holography and $ T\overline{T} $,''
JHEP {\bf 1709}, 136 (2017).
[arXiv:1706.06604 [hep-th]].
}

\lref\CallanDJ{
  C.~G.~Callan, Jr., J.~A.~Harvey and A.~Strominger,
  ``World sheet approach to heterotic instantons and solitons,''
Nucl.\ Phys.\ B {\bf 359}, 611 (1991).
}

\lref\ApoloQPQ{
  L.~Apolo and W.~Song,
  ``Strings on warped AdS$_3$ via $T\bar{J}$ deformations,''
[arXiv:1806.10127 [hep-th]].
}

\lref\DetournayRH{
  S.~Detournay, D.~Israel, J.~M.~Lapan and M.~Romo,
  ``String Theory on Warped $AdS_{3}$ and Virasoro Resonances,''
JHEP {\bf 1101}, 030 (2011).
[arXiv:1007.2781 [hep-th]].
}

\lref\AharonyVK{
  O.~Aharony, B.~Fiol, D.~Kutasov and D.~A.~Sahakyan,
  ``Little string theory and heterotic / type II duality,''
Nucl.\ Phys.\ B {\bf 679}, 3 (2004).
[hep-th/0310197].
}

\lref\ZamolodchikovCE{
  A.~B.~Zamolodchikov,
  ``Expectation value of composite field T anti-T in two-dimensional quantum field theory,''
[hep-th/0401146].
}

\lref\AharonyBAD{
  O.~Aharony, S.~Datta, A.~Giveon, Y.~Jiang and D.~Kutasov,
  ``Modular invariance and uniqueness of $T\bar{T}$ deformed CFT,''
[arXiv:1808.02492 [hep-th]].
}

\lref\AharonyICS{
  O.~Aharony, S.~Datta, A.~Giveon, Y.~Jiang and D.~Kutasov,
  ``Modular covariance and uniqueness of $J\bar{T}$ deformed CFTs,''
[arXiv:1808.08978 [hep-th]].
}

\lref\AharonyUB{
  O.~Aharony, M.~Berkooz, D.~Kutasov and N.~Seiberg,
  ``Linear dilatons, NS five-branes and holography,''
JHEP {\bf 9810}, 004 (1998).
[hep-th/9808149].
}

\lref\WittenYU{
  E.~Witten,
  ``On the conformal field theory of the Higgs branch,''
JHEP {\bf 9707}, 003 (1997).
[hep-th/9707093].
}

\lref\AharonyDP{
  O.~Aharony, M.~Berkooz and E.~Silverstein,
  ``Nonlocal string theories on AdS(3) x S**3 and stable nonsupersymmetric backgrounds,''
Phys.\ Rev.\ D {\bf 65}, 106007 (2002).
[hep-th/0112178].
}

\lref\BerkoozUG{
  M.~Berkooz, A.~Sever and A.~Shomer,
  ``'Double trace' deformations, boundary conditions and space-time singularities,''
JHEP {\bf 0205}, 034 (2002).
[hep-th/0112264].
}

\lref\WittenUA{
  E.~Witten,
  ``Multitrace operators, boundary conditions, and AdS / CFT correspondence,''
[hep-th/0112258].
}

\lref\DubovskyBMO{
  S.~Dubovsky, V.~Gorbenko and G.~Hernández-Chifflet,
  ``$ T\overline{T} $ partition function from topological gravity,''
JHEP {\bf 1809}, 158 (2018).
[arXiv:1805.07386 [hep-th]].
}

\lref\IsraelVV{
  D.~Israel, C.~Kounnas, D.~Orlando and P.~M.~Petropoulos,
  ``Electric/magnetic deformations of S**3 and AdS(3), and geometric cosets,''
Fortsch.\ Phys.\  {\bf 53}, 73 (2005).
[hep-th/0405213].
}

\lref\DetournayRH{
  S.~Detournay, D.~Israel, J.~M.~Lapan and M.~Romo,
  ``String Theory on Warped $AdS_{3}$ and Virasoro Resonances,''
JHEP {\bf 1101}, 030 (2011).
[arXiv:1007.2781 [hep-th]].
}

\lref\DonnellyBEF{
  W.~Donnelly and V.~Shyam,
  ``Entanglement entropy and $T \overline{T}$ deformation,''
Phys.\ Rev.\ Lett.\  {\bf 121}, no. 13, 131602 (2018).
[arXiv:1806.07444 [hep-th]].
}

\lref\ChakrabortyAJI{
  S.~Chakraborty,
  ``Wilson loop in a $T\bar{T}$ like deformed $\rm{CFT}_2$,''
Nucl.\ Phys.\ B {\bf 938}, 605 (2019).
[arXiv:1809.01915 [hep-th]].
}
\lref\BonelliKIK{
  G.~Bonelli, N.~Doroud and M.~Zhu,
  ``$T \bar{T}$-deformations in closed form,''
JHEP {\bf 1806}, 149 (2018).
[arXiv:1804.10967 [hep-th]].
}
\lref\BabaroCMQ{
  J.~P.~Babaro, V.~F.~Foit, G.~Giribet and M.~Leoni,
  ``$ T\overline{T} $ type deformation in the presence of a boundary,''
JHEP {\bf 1808}, 096 (2018).
[arXiv:1806.10713 [hep-th]].
}
\lref\ContiJHO{
  R.~Conti, L.~Iannella, S.~Negro and R.~Tateo,
  ``Generalised Born-Infeld models, Lax operators and the $ T\bar{T} $ perturbation,''
JHEP {\bf 1811}, 007 (2018).
[arXiv:1806.11515 [hep-th]].
}
\lref\ChenEQK{
  B.~Chen, L.~Chen and P.~X.~Hao,
  ``Entanglement entropy in $T\overline{T}$-deformed CFT,''
Phys.\ Rev.\ D {\bf 98}, no. 8, 086025 (2018).
[arXiv:1807.08293 [hep-th]].
}
\lref\GorbenkoOOV{
  V.~Gorbenko, E.~Silverstein and G.~Torroba,
  ``dS/dS and $ T\overline{T} $,''
JHEP {\bf 1903}, 085 (2019).
[arXiv:1811.07965 [hep-th]].
}
\lref\AraujoRHO{
  T.~Araujo, E.~O.~Colgain, Y.~Sakatani, M.~M.~Sheikh-Jabbari and H.~Yavartanoo,
  ``Holographic integration of $T \bar{T}$ $\&$ $J \bar{T}$ via $O(d,d)$,''
JHEP {\bf 1903}, 168 (2019).
[arXiv:1811.03050 [hep-th]].
}
\lref\GuicaVNB{
  M.~Guica,
  ``On correlation functions in $J\bar T$-deformed CFTs,''
J.\ Phys.\ A {\bf 52}, no. 18, 184003 (2019).
[arXiv:1902.01434 [hep-th]].
}
\lref\NakayamaUJT{
  Y.~Nakayama,
  ``Very Special $T\bar{J}$ deformed CFT,''
[arXiv:1811.02173 [hep-th]].
}
\lref\TaylorXCY{
  M.~Taylor,
  ``TT deformations in general dimensions,''
[arXiv:1805.10287 [hep-th]].
}
\lref\HartmanTKW{
  T.~Hartman, J.~Kruthoff, E.~Shaghoulian and A.~Tajdini,
  ``Holography at finite cutoff with a $T^2$ deformation,''
JHEP {\bf 1903}, 004 (2019).
[arXiv:1807.11401 [hep-th]].
}
\lref\CaputaPAM{
  P.~Caputa, S.~Datta and V.~Shyam,
  ``Sphere partition functions and cut-off AdS,''
[arXiv:1902.10893 [hep-th]].
}
\lref\BanerjeeEWU{
  A.~Banerjee, A.~Bhattacharyya and S.~Chakraborty,
  ``Entanglement Entropy for $TT$ deformed CFT in general dimensions,''
[arXiv:1904.00716 [hep-th]].
}
\lref\MurdiaFAX{
  C.~Murdia, Y.~Nomura, P.~Rath and N.~Salzetta,
  ``Comments on Holographic Entanglement Entropy in TT Deformed CFTs,''
[arXiv:1904.04408 [hep-th]].
}
\lref\JiangTCQ{
  Y.~Jiang,
  ``Expectation value of $T\bar{T}$ operator in curved spacetimes,''
[arXiv:1903.07561 [hep-th]].
}
\lref\ContiDXG{
  R.~Conti, S.~Negro and R.~Tateo,
  ``Conserved currents and $T\bar{T}_s$ irrelevant deformations of 2D integrable field theories,''
[arXiv:1904.09141 [hep-th]].
}
\lref\CardyJHO{
  J.~Cardy,
  ``$T\overline T$ deformations of non-Lorentz invariant field theories,''
[arXiv:1809.07849 [hep-th]].
}
\lref\ContiTCA{
  R.~Conti, S.~Negro and R.~Tateo,
  ``The $T\bar{T} $ perturbation and its geometric interpretation,''
JHEP {\bf 1902}, 085 (2019).
[arXiv:1809.09593 [hep-th]].
}
\lref\ChangDGE{
  C.~K.~Chang, C.~Ferko and S.~Sethi,
  ``Supersymmetry and $T \overline{T}$ Deformations,''
[arXiv:1811.01895 [hep-th]].
}
\lref\JiangHUX{
  H.~Jiang, A.~Sfondrini and G.~Tartaglino-Mazzucchelli,
  ``$T\bar{T}$ deformations with $\NN=(0,2)$ supersymmetry,''
[arXiv:1904.04760 [hep-th]].
}
\lref\BaggioRPV{
  M.~Baggio, A.~Sfondrini, G.~Tartaglino-Mazzucchelli and H.~Walsh,
  ``On $T\bar{T}$ deformations and supersymmetry,''
[arXiv:1811.00533 [hep-th]].
}

\lref\GuicaMU{
  M.~Guica, T.~Hartman, W.~Song and A.~Strominger,
  ``The Kerr/CFT Correspondence,''
Phys.\ Rev.\ D {\bf 80}, 124008 (2009).
[arXiv:0809.4266 [hep-th]].
}

\lref\GiveonJJ{
  A.~Giveon and M.~Rocek,
  ``Generalized duality in curved string backgrounds,''
Nucl.\ Phys.\ B {\bf 380}, 128 (1992).
[hep-th/9112070].
}

\lref\GiveonPH{
  A.~Giveon and E.~Kiritsis,
  ``Axial vector duality as a gauge symmetry and topology change in string theory,''
Nucl.\ Phys.\ B {\bf 411}, 487 (1994).
[hep-th/9303016].
}

\lref\SonYE{
  D.~T.~Son,
  ``Toward an AdS/cold atoms correspondence: A Geometric realization of the Schrodinger symmetry,''
Phys.\ Rev.\ D {\bf 78}, 046003 (2008).
[arXiv:0804.3972 [hep-th]].
}

\lref\BalasubramanianDM{
  K.~Balasubramanian and J.~McGreevy,
  ``Gravity duals for non-relativistic CFTs,''
Phys.\ Rev.\ Lett.\  {\bf 101}, 061601 (2008).
[arXiv:0804.4053 [hep-th]].
}

\lref\AzeyanagiZD{
  T.~Azeyanagi, D.~M.~Hofman, W.~Song and A.~Strominger,
  ``The Spectrum of Strings on Warped $AdS_3\times S^3$,''
JHEP {\bf 1304}, 078 (2013).
[arXiv:1207.5050 [hep-th]].
}

\Title{} {\centerline{$T\bar{T}$, $J\bar{T}$, $T\bar{J}$ and String Theory}}

\bigskip
\centerline{\it Soumangsu Chakraborty${}^{1}$, Amit Giveon${}^{1}$ and David Kutasov${}^{2}$}
\bigskip
\smallskip
\centerline{${}^{1}$Racah Institute of Physics, The Hebrew
University} \centerline{Jerusalem 91904, Israel}
\smallskip
\centerline{${}^2$EFI and Department of Physics, University of
Chicago} \centerline{5640 S. Ellis Av., Chicago, IL 60637, USA }

\smallskip

\vglue .3cm

\bigskip

\bigskip
\noindent
We calculate the spectrum of a two dimensional CFT on a cylinder, perturbed by a general linear combination of $T\bar{T}$, $J\bar{T}$ and $T\bar{J}$, by utilizing the relation of this problem to certain solvable single trace current-current deformations of the worldsheet theory of strings on $AdS_3$. We show that this spectrum is well behaved (\ie\ the energies are real) if and only if the dual bulk geometry is well behaved (\ie\ the signature of the bulk spacetime is $(1,2)$ and there are no closed timelike curves). We also comment on the relations of string theory on $AdS_3$ and its deformations to symmetric product CFT's, matrix string theory, and a black hole description of highly excited fundamental strings.

\bigskip

\bigskip

\bigskip

\noindent
{\it Dedicated to the memory of P. G. O. Freund}

\Date{}

\newsec{Introduction}

About three years ago, two groups of researchers \refs{\SmirnovLQW,\CavagliaODA} discovered a class of irrelevant deformations of two dimensional Quantum Field Theories (QFT's) that, unlike generic such deformations, appear to be under control at finite values of the coupling. The resulting theories, often referred to as $T\bar T$ deformed QFT's, or $T\bar T$ for short, are of interest from a number of perspectives.

From the general QFT point of view, they are interesting because of the lore that theories with irrelevant deformations are always effective, \ie\ have a built in physical cutoff. In $T\bar T$, one can apparently flow up the RG without encountering ambiguities and/or singularities.\foot{For one sign of the coupling, that we will refer to as positive below. For negative coupling one encounters both ambiguities and singularities.} The mechanism for this is a prescription for flowing up the RG by an infinitesimal amount, which involves a result of A. Zamolodchikov from 2004 \ZamolodchikovCE.

Another reason for the interest in these theories is that they do not approach a conformal field theory (CFT) at high energies. Rather, for positive coupling, their high energy spectrum exhibits Hagedorn growth, and appears to give the first example of a theory with this behavior which is under analytic control. For negative coupling, one finds that highly excited states have complex energies. Thus, the theory appears to be non-unitary in this case.

$T\bar T$ deformed QFT is also of interest in the study of strings in confining gauge theories, and in particular scattering on the worldsheet of such strings. It was discussed from that point of view in \refs{\DubovskyWK\DubovskyIRA-\DubovskyCNJ} and other papers. It also seems to be related to a theory of gravity (known as JT gravity) \refs{\DubovskyCNJ,\DubovskyBMO}, and to the dynamics on (non-)critical fundamental strings wrapped around a circle \refs{\CavagliaODA}.

In one of the initial papers on the subject,  \SmirnovLQW, it was pointed out that the $T\bar T$ construction can be generalized to a large class of theories which contain conserved currents, with the analog of $T\bar T$ given by a product of currents. An example of such a theory was proposed in \GuicaLIA\ and solved in \ChakrabortyVJA. In these papers, the unperturbed theory was taken to be a CFT with a left-moving conserved current $J(x)$, and the analog of $T\bar T$ is the product $J\bar T$.

The resulting theory, which we will refer to as $J\bar T$ deformed CFT, has the property that the energies of highly excited states in it are complex for all values of the coupling \ChakrabortyVJA. Thus, at least naively, it is not unitary. A possible solution to this problem is to consider perturbations of two dimensional QFT's that involve a linear combination of $T\bar T$  and $J\bar T$. Assuming that such perturbations  lead to a sensible theory, one may hope that for a sufficiently large $T\bar T$ coupling, unitarity is restored. This provides a strong motivation to study theories that involve both types of perturbations.

As is well known, many two dimensional CFT's are related by holography to string theory on $AdS_3$. It is natural to ask what is the effect of deformations such as $T\bar T$, $J\bar T$, etc, in this context. Operators like $J$, $T$ and $\bar T$ correspond in the bulk string theory to ``single trace operators,'' \ie\ vertex operators integrated over the worldsheet of the string. Thus, operators such as $T\bar T$ and $J\bar T$ correspond to double trace operators. The effect of adding such operators to the spacetime Lagrangian in string theory was studied in \refs{\AharonyDP\WittenUA-\BerkoozUG}. To leading order in the coupling, it corresponds to modifying the boundary conditions for the graviton on $AdS_3$; the finite perturbation is harder to study using bulk techniques.

An intriguing holographic dual of $T\bar T$ deformed CFT with negative coupling was proposed in \McGoughLOL. It involves string theory on $AdS_3$ with a physical UV cutoff at a radial location related to the value of the coupling. This proposal was further discussed in \refs{\ShyamZNQ\KrausXRN\CottrellSKZ-\DonnellyBEF}. Some questions about it were raised, and its status is at present unclear. For positive coupling, which corresponds to a more well behaved theory (in particular, one in which all energies are real), there is no concrete proposal of a holographic dual. Thus, holography does not seem to shed much light on questions like what is the bulk interpretation of the exact solvability of these theories, of the Hagedorn spectrum one finds for positive coupling, etc.

Interestingly, string theory contains a class of operators that are closely related to $T\bar T$, $J\bar T$, etc, that lead to deformations that can be studied exactly in the full string theory on $AdS_3$ (with Neveu-Schwartz $H$-flux) \GiveonNIE. Unlike $T\bar T$ and its cousins, they are single trace operators. Thus, the associated deformations correspond to adding local operators to the worldsheet Lagrangian. As we review below, these operators are products of left and right-moving currents on the worldsheet, and give rise to an exactly solvable theory -- a generalization of the abelian Thirring model.

The interplay between the study of these deformations in string theory and field theory has been a fruitful direction of research in the last two years. It sheds light on both sides of the duality; the string approach allows one to calculate things that are hard to calculate in field theory, while the field theory leads to an improved understanding of holography in spacetimes that are not asymptotically AdS (\eg\ ones obtained by climbing out of the near-horizon region of fundamental strings).

Some special cases of the construction were discussed in previous papers. In particular, $T\bar T$ deformed CFT was studied in \refs{\GiveonNIE\GiveonMYJ\AsratTZD\GiribetIMM\ChakrabortyKPR-\ChakrabortyAJI}, and $J\bar T$ in \refs{\ChakrabortyVJA,\ApoloQPQ,\GiveonFGR}. It was shown in these papers that the string theory construction gives the same spectrum as the field theory, and the presence (absence) of states with complex energies is related to the presence (absence) of pathologies in the dual geometry, in particular, the presence of closed timelike curves (CTC's).

The purpose of this paper is to extend the construction to a general linear combination of $T\bar T$, $J\bar T$ and $T\bar J$ deformations of a CFT which contains a conserved left (right)-moving current $J$ $(\bar J)$. From the string theory point of view, it is clear that the solvability of the model should extend to all values of the couplings, since on the worldsheet it corresponds to a generalized abelian Thirring model. Thus, string theory allows one to compute the spectrum of a field theory that is harder to study directly.

One of the goals of the study is to expand the set of theories for which the spectrum is real, and analyze their properties as a function of the couplings. Another goal is to compare the conditions on the couplings that are necessary for the spectrum to be real to the conditions that are necessary for the dual bulk geometry to be regular.

The plan of the paper is the following. In section 2, we review some features of string theory on $AdS_3$ that play a role in our later discussion. We discuss the subsector of the theory that describes the dynamics of long strings, and its description as a symmetric product CFT. We also discuss some of the evidence for the conjecture that the full string theory on $AdS_3$ is described by a symmetric product CFT, and some properties that the block CFT must have in that case.

In section 3, we introduce the irrelevant single trace deformations of interest. We describe them as current-current deformations of the worldsheet CFT, and as deformations of the building block of the symmetric product CFT describing long strings.

In section 4, we study the spacetime geometry of the deformed theory. We write the worldsheet sigma model action and read off from it the spacetime fields that form this background. We find the conditions on the couplings that  are necessary for the background to be regular, have the usual Lorentzian signature, and no CTC's.

In section 5, we compute the string spectrum in the deformed background. We read off from it the spectrum of a general CFT perturbed by a general combination of $T\bar T$, $J\bar T$ and $T\bar J$, and study the conditions on the couplings that are necessary for all energies to be real. We find that these conditions are exactly the same as those found in section 4 for the regularity of the dual background.

In section 6, we summarize and discuss our results. Four appendices contain some details,
concerning results and statements used in the text.

For some additional interesting work on related subjects see \refs{\CardySDV\AharonyVUX\BzowskiPCY\BonelliKIK\TaylorXCY\DattaTHY\BabaroCMQ\ContiJHO\ChenEQK\HartmanTKW\AharonyBAD\AharonyICS\CardyJHO\ContiTCA\BaggioRPV\ChangDGE\NakayamaUJT\AraujoRHO\GorbenkoOOV\GuicaVNB\CaputaPAM\JiangTCQ\BanerjeeEWU\MurdiaFAX\JiangHUX-\ContiDXG}.
\smallskip
{\it Note added:} After this work was completed, we received \refs{\LeFlochRUT\AharonyBAD}, which studies the spectrum of the theory with all couplings turned on. Their results appear to agree with ours.

\newsec{Aspects of string theory on $AdS_3$}

In this section we briefly discuss perturbative string theory on $AdS_3$ with NS-NS H-flux, as preparation for the study of a class of irrelevant deformations of this theory in later sections. The understanding of this theory is incomplete at present, but we will describe some things that are known about it, and mention some that are more conjectural. When discussing examples, we will mainly consider backgrounds that preserve $\NN=2$ or larger supersymmetry, such as $AdS_3\times S^3\times T^4$ (which preserves $(4,4)$ SUSY).

The worldsheet theory for a string propagating on $AdS_3$ is described by a $\sigma$-model on the $SL(2,\IR)$ group manifold. It is invariant under left and right-moving $SL(2,\IR)$ current algebras at level $k$. This level is related to the radius of curvature of $AdS_3$ in string units, $R_{\rm AdS}=\sqrt{k} l_s$. The current algebra  plays an important role in analyzing the theory, and in particular in studying its spectrum, symmetries and correlation functions. In particular, the zero modes of the left-moving currents $J_{\rm{SL}}^-(z)$, $J_{\rm{SL}}^3(z)$ and $J_{\rm{SL}}^+(z)$ give in spacetime the left-moving global conformal charges $L_{-1}$, $L_0$ and $L_1$, respectively, and similarly for the right-movers. The full Virasoro algebra in spacetime is also constructed using these currents \refs{\GiveonNS,\KutasovXU}.

The AdS/CFT correspondence relates string theory on $AdS_3$ to a two-dimensional conformal field theory living on the boundary, which is usually  refered to as the boundary or spacetime CFT. For pure NS-NS H-flux, this theory has the following properties, that will be important for our discussion below:
\item{(1)} The theory has an $SL(2,\IC)$ invariant vacuum,\foot{In Lorentzian signature, $SL(2,\IC)$ is replaced by $SL(2,\IR)\times SL(2,\IR)$.} the NS vacuum, which corresponds in the bulk description to $AdS_3$ in global coordinates. The boundary of global $AdS_3$ is a cylinder, with SUSY broken by boundary conditions. The Ramond vacuum of the spacetime CFT, which preserves SUSY, corresponds in the bulk description to the $M=J=0$ BTZ black hole.

\item{(2)} In the NS sector, the spectrum of excitations consists of a sequence of discrete states which come from the principal discrete series representations of $SL(2,\IR)$, followed by a continuum of long string states. The continuum starts at dimension $\sim k/2$, and is described by applying spectral flow to the principal continuous series states. There are also discrete states obtained by applying spectral flow to the principal discrete series \MaldacenaHW.

\item{(3)} In the Ramond sector, the spectrum contains a continuum of long string states above a gap of order $1/k$, associated with the principal continuous series. It may also have states coming from the principal discrete series.

\item{(4)} The theory on a single long string in global $AdS_3$ was analyzed in \SeibergXZ. In string theory on $AdS_3\times\NN$, it is described by a $\sigma$-model on\foot{The superscript $(L)$ stands for long string. The subscript will be explained momentarily.}
\eqn\aaa{\MM_{6k}^{(L)}=\IR_\phi\times \NN.}
The coordinate $\phi$ parametrizes the location of the string in the radial direction of $AdS_3$; the boundary is at $\phi\to\infty$. The theory on $\IR_\phi$ has a linear dilaton with slope
\eqn\bbb{Q^{(L)}=(k-1)\sqrt{2\over k}~.}
The central charge of the theory \aaa\ is given by
 \eqn\ccc{c_\MM=6k,}
 which explains the subscript on the l.h.s. of \aaa. An example is string theory on $AdS_3\times S^3\times T^4$, where $\MM_{6k}^{(L)}=\IR_\phi\times SU(2)_k\times T^4$, a CFT with $(4,4)$ superconformal symmetry.
\item{(5)} The effective string coupling on the long string, $g_s\sim \exp(Q^{(L)}\phi)$, increases as the string moves towards the boundary.\foot{For $k>1$; for $k<1$ it decreases, and the physics is different \GiveonMI.} Thus, the physics of long strings near the boundary is strongly coupled. This observation plays an important role in studying the spacetime CFT \GiveonMI.

\noindent
An interesting open problem is what is the spacetime CFT corresponding to a given $AdS_3$ vacuum. There is strong evidence for the conjecture that the long string states are well described by  the symmetric product CFT
\eqn\ddd{\left(\MM_{6k}^{(L)}\right)^p/S_p~,}
where $p$ is the number of strings that make up the vacuum.\foot{One can think of $AdS_3$ backgrounds of the sort discussed here as obtained from linear dilaton backgrounds, \ie\ vacua of Little String Theory \refs{\CallanDJ,\AharonyUB}, by adding to them $p$ fundamental strings \GiveonZM. This gives rise in the near-horizon geometry of the strings to an $AdS_3$ vacuum with string coupling $g_s^2\sim 1/p$.}
One way to understand this is to follow the logic of matrix string theory \refs{\MotlTH,\DijkgraafVV}. It was shown in that context, that if the theory living on a string wrapping once around a circle is $\MM$, then the symmetric product CFT $\MM^N/S_N$ provides an effective description of the Hilbert space of $N$ free strings. Untwisted sector states describe $N$ strings, each winding once around the circle, while $Z_w$ twisted states (with $w=2,3,\cdots, N$) describe strings that wind $w$ times around the circle. General states corresponding to $n$ strings with  windings $(w_1,\cdots, w_n)$, with $w_1+\cdots+w_n=N$, are described in terms of conjugacy classes of the symmetric group \ddd.

The work on matrix string theory is relevant for long strings on $AdS_3$ since they are weakly coupled in a wide range of positions in the radial direction. As mentioned in (5) above, eventually, when the strings move far enough towards the boundary in the radial direction, their coupling becomes large, and one expects corrections to \ddd\ to become important. We will discuss such corrections below.

Another piece of evidence for the symmetric product structure \ddd\ is the spectrum of fundamental strings on $AdS_3$. As an example, in the Ramond sector of the spacetime CFT, which as mentioned above corresponds to the massless BTZ background in string theory, the spectrum of long strings is given by \ParsonsSI\
\eqn\eee{E_{L,R}={1\over w}\left[-{j(j+1)\over k}+N_{L,R}-\half\right],}
where
\eqn\fff{E_L={R\over2}\left(E+P\right);\;\;\;E_R={R\over2}\left(E-P\right);\;\;\;P\in {1\over R}Z~,}
and $R$ is the radius of the spatial circle on the boundary. $N_{L,R}$ are the left and right-moving excitation levels of the string.

For long strings moving with momentum $p$ in the radial direction one has
\eqn\ggg{j=-\half+is;\;\;\; s\in \IR,}
with $p=s\sqrt{2\over k}$ in string units.
The quantum number $w=1,2,\cdots$ in \eee\ is the winding number of the string around the boundary circle. This quantum number is not conserved.

To make contact with \ddd, we note that in a symmetric product CFT, $\MM^N/S_N$, states  in the $Z_w$ twisted sector have energies
\eqn\hhh{E_L=h_w-{kw\over4};\;\;\; E_R=\bar h_w-{kw\over4}~,
}
where $h_w$, $\bar h_w$ are the left and right-moving scaling dimensions of the operator that creates the state in question from the $SL(2,\IC)$ invariant vacuum. States with $w=1$ are those of the original CFT $\MM$. For every such state with left-moving dimension $h_1$, there is a state in the $Z_w$ twisted sector, with dimension $h_w$ given by \refs{\KlemmDF\FuchsVU - \ArgurioTB}
\eqn\iii{h_w={h_1\over w}+{k\over4}\left(w-{1\over w}\right),}
where we took the central charge of $\MM$ to be $c_\MM=6k$, as in \ddd. Similar formulae apply to the right-movers.

The string spectrum \eee\ has similar properties. A state in the winding one sector $(w=1)$ is labeled by the choice of excitation levels  $N_{L,R}$ and, for a long string, a choice of the radial momentum  $s$, \ggg. For a given value of these parameters,
one can check using
\eee\ -- \hhh\ that \iii\ is satisfied for all states, in agreement with the proposal \ddd, \refs{\GiveonMI,\ArgurioTB}.

There are other motivations for the validity of \ddd\ for long strings, one of which is the study of irrelevant deformations such as $T\bar T$, $J\bar T$, etc. As we discuss in the next sections, assuming \ddd\ leads to non-trivial agreement between field theoretic and string theoretic analyses of the spectra of such irrelevant deformations of CFT's \refs{\GiveonNIE\GiveonMYJ\ChakrabortyVJA-\ApoloQPQ}.

So far we studied long strings on $AdS_3$, and in particular some of the evidence that their dynamics is described by the symmetric product CFT \ddd. This correspondence is sufficient for the analysis in the rest of this paper, and in principle we could stop this section at this point. However, it is natural to ask what is the structure of the full spacetime CFT, and in particular whether it could also take a symmetric product form,
\eqn\jjj{\left(\MM_{6k}\right)^p/S_p~,}
perhaps only at large $p$. In the rest of this section we examine this possibility. We want to stress that we are not claiming that the full spacetime CFT takes the form \jjj. Rather, we are discussing the question what properties the block CFT $\MM_{6k}$ must have, assuming \jjj.

The long string effective theory \ddd\ has the property that the $SL(2,\IC)$ invariant vacuum is not in the spectrum. In each factor $\MM_{6k}^{(L)}$ there is a gap of size $h_{\rm min}={1\over 8}\left(Q^{(L)}\right)^2$ in the spectrum of scaling dimensions. Thus, the gap in the full theory is ${p\over 8}\left(Q^{(L)}\right)^2$. On the other hand, as mentioned above, string theory on $AdS_3$ does have an $SL(2,\IC)$ invariant vacuum. Thus, this must be a property of the block $\MM_{6k}$ in \jjj.

String theory on $AdS_3$ is also believed to give rise to a unitary and modular invariant spacetime CFT with $c=6kp$. This, together with the fact that the $SL(2,\IC)$ invariant vacuum is a normalizable state in the theory, implies that the entropy grows at large scaling dimensions, $h,\bar h\gg kp/4$, like
\eqn\kkk{S=2\pi\sqrt{kp}\left(\sqrt{h}+\sqrt{\bar h}\right).}
On the other hand, the long string theory \ddd\ has a smaller entropy, obtained from \kkk\ by replacing $k\to 2-{1\over k}$. Therefore, most of the states contributing to the entropy (for $k>1$, \GiveonMI) are not fundamental string states. This is not surprising, as in general the Cardy entropy \kkk\ agrees with the Bekenstein-Hawking entropy of black holes \StromingerEQ, in this case BTZ black holes in $AdS_3$.  Thus, one expects most of the entropy to come from non-perturbative states in string theory, and to be larger than that of perturbative strings.

What is somewhat surprising is that if the symmetric product picture \jjj\ is correct, the energy at which such non-perturbative states first appear is lower than what one might expect. Indeed, consider the block generating the symmetric product \jjj, $\MM_{6k}$. Since, as mentioned above, it is expected to be unitary, modular invariant and contain an $SL(2,\IC)$ invariant vacuum, its entropy must approach \kkk\ with $p=1$ at large scaling dimensions, $h,\bar h\gg k/4(=c_\MM/24)$. On the other hand, as mentioned above, the spectrum of perturbative strings has the same growth with $k\to 2-{1\over k}$, which is smaller (for $k>1$). Thus, most of the states with $h,\bar h\gg k/4$ are non-perturbative.

This is surprising since it means that the perturbative string description of this theory breaks down at energies that remain finite in the limit $g_s\to 0$ (recall that in this class of theories $g_s^2\sim 1/p$). Usually in string theory, physics at energy scales that remain finite in the limit $g_s\to 0$ is well described by the worldsheet analysis.

One way to understand this phenomenon is to think about the physics from the point of view of the Little String Theory (LST) underlying the $AdS_3$ vacuum in question. As mentioned above, we can think of string theory on $AdS_3\times\NN$ as obtained from a two dimensional vacuum of LST, described in the bulk  by the linear dilaton background $\IR_\phi\times\IR_t\times S^1\times\NN$ \refs{\CallanDJ\AharonyUB-\GiveonZM}, by placing $p$ fundamental strings wrapping the $S^1$ into the throat labeled by $\phi$. A long string state (say with $w=1$) of the sort described above corresponds to a configuration where $p-1$ strings condense at $\phi\to-\infty$ while the last one is separated from them in the radial direction. When this separation is large, the geometry near that string looks like $AdS_3\times\NN$, but now with $p=1$, \ie\ the coupling near the string is not small. Hence, the theory on the string, which perturbatively is  $\MM_{6k}^{(L)}$ \aaa, has many more states obtained by replacing the $AdS_3$ geometry with $p=1$ in a state with a perturbative string excitation, by a BTZ black hole with the same quantum numbers.

This explains why the perturbative description of the theory on a long string is corrected due to non-perturbative effects, in the regime where long strings can exist. In the NS sector of the spacetime CFT, this happens for scaling dimensions above the gap to the long strings, $h=k/4$. In the Ramond sector, the gap is of order $1/k$ (see appendix A for a more detailed discussion).\foot{Note that this does not invalidate the discussion of perturbative long string states around equations \eee\ -- \ggg, since the wavefunctions of such states have support in the wide region where the theory is weakly coupled. Rather, it implies that there are additional states that are supported in the strongly coupled region, that have energies comparable to those of the perturbative states. The situation is similar to that in Liouville-type theories, where there is a continuum of scattering states living far from the wall, and there may be additional states whose wavefunctions are localized near the wall, some or all of which may be non-perturbative.}

The above discussion clarifies the sense in which the spacetime CFT corresponding to string theory with purely NS-NS H-flux is singular \SeibergXZ. In addition to the continuum above a gap, it has the property that non-perturbative (in $g_s$) effects affect the physics at energy scales that do not depend on the value of the underlying string coupling, which can be taken to be arbitrarily small. The situation is reminiscent of string theory on a singular or near-singular Calabi-Yau manifold, where the underlying string coupling can be taken to be arbitrarily small, but the masses of non-perturbative states such as localized D-branes can be held fixed, and even go to zero, as one approaches a singular point in moduli space (see \eg\ \refs{\StromingerCZ,\GiveonZM,\GiveonPX,\AharonyVK}).

The discussion above focused on the relation of the perturbative block $\MM_{6k}^{(L)}$ \aaa\ to the non-perturbative one $\MM_{6k}$ \jjj. In matrix string theory, the description of wrapped strings also involves turning on a twist field that exchanges copies of the block \refs{\MotlTH,\DijkgraafVV}. In our case such a twist field is irrelevant for $k>1$, and hence cannot appear in the low energy theory.\foot{In fact, this observation provides an additional motivation for the conjecture that the full spacetime CFT corresponding to string theory on $AdS_3$ takes the symmetric product form \jjj.}
For $k=1$, it is marginal and thus can play a role; in general, one expects it to.

We finish this section with a few comments:
\item{(1)} The usual BTZ black holes dominate the spectrum for dimensions $h\gg kp/4(=c/24)$. In the description \jjj, one can think of them as states in which the energy is distributed among all the factors in the symmetric product, $E=E_1+E_2+\cdots+E_p$. Most of the entropy is in states where the energy is divided roughly equally among all the factors.
\item{(2)} An example of the above discussion is string theory on $AdS_3\times S^3\times T^4$. A natural candidate for the block $\MM_{6k}$ in this case is\foot{The discussion below is related to that of \WittenYU.} $\MM_{6k}=\left(T^4\right)^k/S_k$. For $k\ge 2$ this model has a modulus in the $Z_2$ twisted sector, and for a particular value of this modulus the CFT is singular; this is the point corresponding to $\MM_{6k}$. The region near the singularity is described by a throat theory of the sort \aaa. Many of its properties can be determined by symmetry considerations. The CFT $\left(T^4\right)^k/S_k$ has $(4,4)$ superconformal symmetry and central charge $6k$. The superconformal algebra includes an $SU(2)_R$ current algebra at level $k$. This algebra must be realized in the throat theory as well. This, together with the expected appearance of a linear dilaton direction, $\phi$, leads to the expectation that the throat of this theory contains a factor $\IR_\phi\times SU(2)_k$. In addition, the theory must contain left and right-moving $U(1)^4$ currents. Thus, the minimal scenario for the form of the throat theory of $\left(T^4\right)^k/S_k$ is $\IR_\phi\times SU(2)_k\times T^4$, with the dilaton slope given by \bbb. This is the same as \aaa, with $\NN=SU(2)_k\times T^4$.
\item{(3)} The case $k=1$ in the discussion above, which corresponds to the low-energy dynamics of $p$ strings near a single fivebrane, is special. Setting $k=1$ in the formulae above, gives $\left(T^4\right)^p/S_p$ (for the spacetime CFT). This theory again contains a modulus in the $Z_2$ twisted sector, but now it is not part of the theory describing the block $\MM_{6k}$. There are two natural values of this modulus: the one corresponding to the standard orbifold theory, and the one corresponding to a singular theory. The discussion above, as well as that of \WittenYU, suggest that the correct value is the singular one. Note that: (a) For $k=1$, the description as string theory on $AdS_3\times S^3\times T^4$ does not make sense, since it involves a negative level $SU(2)$ WZW model \CallanDJ, but the infrared theory of the brane system above should still exist; (b) U-duality relates the case of $p$ strings and one fivebrane to that of $p$ fivebranes and one string, which was discussed above. The CFT is the same in both cases, and it is singular.

\newsec{Some solvable irrelevant deformations of string theory on $AdS_3$}

As described in the previous section, string theory on $AdS_3$ is equivalent to a two-dimensional CFT. Thus, one can use it to study irrelevant deformations of the sort discussed in the last few years, such as $T\bar T$ \refs{\SmirnovLQW,\CavagliaODA} and $J\bar T$ \refs{\GuicaLIA,\ChakrabortyVJA}. From the bulk point of view, these are multi-trace deformations. Interestingly, the understanding of these deformations in string theory is much less complete than in field theory. In particular, there is no understanding from the string theory point of view why these deformations are exactly solvable in field theory.

At the same time, string theory naturally contains a class of closely related deformations, that are much better understood \refs{\GiveonNIE,\GiveonMYJ,\ChakrabortyVJA,\ApoloQPQ}. In this section we briefly review these deformations, as preparation for the analysis of the next few sections.

String theory in any background of the form $AdS_3\times\NN$ contains an operator, $D(x,\bar x)$,\foot{$(x,\bar x)$ are coordinates on the boundary of $AdS_3$.} constructed in \KutasovXU, that has many properties in common with the operator $T(x)\bar T(\bar x)$ in the spacetime CFT.  This operator is a quasi-primary of dimension $(2,2)$, and has the same OPE with the stress-tensor as $T\bar T$. However, unlike $T\bar T$, $D$ is a single trace operator, which is described in string theory by a vertex operator integrated over the worldsheet, rather than products of such integrated vertex operators, which correspond to multi-trace operators.

It is natural to ask what happens when we perturb the spacetime Lagrangian by $D(x,\bar x)$. As shown in \GiveonNIE, this is the same as adding to the worldsheet Lagrangian the operator
\eqn\lll{\delta\CL=\lambda J_{\rm{SL}}^-\bar {J}_{\rm{SL}}^-~,}
where $J_{\rm{SL}}^-$ $(\bar{J}_{\rm{SL}}^-)$ is the left-moving (right-moving) $SL(2,\IR)$ current whose zero mode gives rise to the spacetime Virasoro generator $L_{-1}$ $(\bar L_{-1})$. Already at this level, we see a number of parallels between the deformation \lll\ and $T\bar T$ in the spacetime CFT. In particular, since the currents $J_{\rm{SL}}^-$ and $\bar{J}_{\rm{SL}}^-$ have spacetime scaling dimensions $(1,0)$ and $(0,1)$, respectively, $\lambda$ in \lll\ has dimension $(-1,-1)$, like the $T\bar T$ coupling. Furthermore, the deformation \lll\ is exactly solvable in string theory, since it corresponds to a truly marginal deformation of the worldsheet theory.  This is reminiscent of the fact that the $T\bar T$ deformation is exactly solvable in field theory.

To better understand the relation between \lll\ and $T\bar T$, we appeal to the discussion of the previous section. In particular, we would like to understand the action of this deformation on the long string states. As we saw in section 2, these states are described by the symmetric product CFT \ddd. Thus, we need to understand the role of the coupling $\lambda$ in that CFT.

Local single-trace operators in string theory correspond in the symmetric product to operators of the form
\eqn\mmm{\sum_{j=1}^p\OO_j(x,\bar x),}
where $\OO_j$ is an operator in the $j$'th copy of the block of the symmetric product. Thus, we expect $D(x,\bar x)$ to be expressible as
\eqn\nnn{D(x,\bar x)=\sum_{j=1}^pD_j(x,\bar x),}
where $D_j$ is a dimension $(2,2)$ quasi-primary operator living in the $j$'th copy of $\MM_{6k}^{(L)}$. There is a natural candidate for such an operator, $D_j=T_j\bar T_j$, the product of the holomorphic and anti-holomorphic stress-tensors in the $j$'th block. In a particular $AdS_3$ vacuum there could be additional operators with the right properties, however, the fact that the operator $D(x,\bar x)$ is universal -- it exists in all $AdS_3$ vacua of string theory -- implies that it is proportional to a sum of $T\bar T$ operators in each block. Indeed, for general $\NN$, string theory on $AdS_3\times\NN$ does not contain another operator with the same properties. Therefore,  $D_j$ \nnn\ must be identified with $T_j\bar T_j$.

If the spacetime CFT is supersymmetric, with $(2,2)$ or larger supersymmetry, we have some additional structure. In the bulk, vacua with $(2,2)$ SUSY correspond to backgrounds of the form $AdS_3\times\NN$, where $\NN$ contains $U(1)_L$ and $U(1)_R$ current algebras (on the worldsheet), and $\NN/U(1)$ is $(2,2)$ superconformal \refs{\GiveonJG,\BerensteinGJ}.
In such theories, the operator $D(x,\bar x)$ is a top component of a spacetime superfield, whose bottom component has dimension $(1,1)$. This dimension $(1,1)$ operator has the same OPE's with the (super) currents as the operator $\sum_{i=1}^p J_i\bar J_i$, where $J_i$ $(\bar J_i)$ is the left (right)-moving $U(1)_R$ current in the spacetime SCFT $\MM_{6k}^{(L)}$, and $i$ labels the copies.

Thus, the question of uniqueness of the identification of $D(x,\bar x)$ with $\sum_jT_j\bar T_j$ reduces in this case to the question whether in a $(2,2)$ superconformal field theory there exists a dimension $(1,1)$ operator with the same OPE's with the currents as $J(x)\bar J(\bar x)$. In general the answer is no, and since the string construction applies to a large class of models (different $\NN$'s with the properties listed above), this identification must be correct.

We conclude that turning on the coupling \lll\ in string theory on $AdS_3$ corresponds, in the symmetric product \ddd\ that describes long strings, to a $T\bar T$ deformation of the block generating the symmetric product $\MM_{6k}^{(L)}$. In particular, the energies of long string states, \eee\ -- \ggg, must evolve under the deformation \lll\ in the same way as they would under the dual $T\bar T$ deformation. It has been shown in \refs{\GiveonNIE,\GiveonMYJ,\ChakrabortyVJA} that this is indeed the case. This provides strong evidence for the picture presented in section 2, and gives a powerful tool for studying $T\bar T$ deformed CFT and related theories using string theory.

The string description sheds light on certain qualitative properties of the spectrum of $T\bar T$ deformed CFT. The original analysis of \refs{\SmirnovLQW,\CavagliaODA} found that the spectrum depends sensitively on the sign of the coupling. For positive coupling one finds a real spectrum with Hagedorn growth, while for negative coupling the spectrum is pathological -- the energies of highly excited states in the original CFT become complex after the deformation, which seems inconsistent with unitarity.\foot{For positive coupling, the NS ground state energy turns complex if the coupling is larger than a certain critical value. As explained in \AharonyBAD, this can be understood as follows. $T\bar T$ deformed CFT is a non-local theory, as reflected in its Hagedorn spectrum. Compactifying it on a circle whose size is below the non-locality scale leads to the above phenomenon. This is qualitatively different from what happens for the other sign of the coupling, for which energies of highly excited states are complex for all values of $R$.}

In the string description, these properties reflect the geometry of the deformed spacetime corresponding to \lll. For positive coupling, the deformation \lll\ modifies the $AdS_3$ spacetime in the UV (near the boundary) to an asymptotically linear dilaton one,
which describes a two-dimensional vacuum of Little String Theory. The latter is known to have a Hagedorn density of states (see \eg\ \KutasovJP),
which explains the appearance of this behavior in the deformed theory \lll.

For negative coupling, the deformation \lll\ modifies the $AdS_3$ in the UV to a geometry which has a singularity at a finite value of the radial coordinate, beyond which one finds closed timelike curves. This explains the pathological UV behavior of the spectrum -- it results from string quantization in a pathological spacetime. It has been argued in \refs{\AharonyICS,\GiveonFGR} that the breakdown of unitarity is associated with the closed timelike curves (rather than the singularity);  we will  provide more support for this below.

The above discussion has an interesting generalization that involves affine Lie algebras. If the internal worldsheet CFT $\NN$ contains left (right)-moving conserved currents $K(z)$ $(\bar K(\bar z))$, one can construct left (right)-moving currents in the spacetime theory $J(x)$
$(\bar J(\bar x))$ \KutasovXU. Following the construction of $D(x,\bar x)$, one can also construct dimension $(1,2)$ and $(2,1)$ operators, $A(x,\bar x)$ and $\bar A(x,\bar x)$, respectively \KutasovXU. These operators have the following properties:
\item{(1)} $A(x,\bar x)$ has the same dimension and OPE's with the currents as $J(x)\bar T(\bar x)$. However, unlike that operator, it is a single trace operator. $\bar A(x,\bar x)$ is related in a similar way to $T(x)\bar J(\bar x)$.
\item{(2)} In the symmetric product CFT \ddd, one can think of the operator $A$ as
\eqn\ooo{A(x,\bar x)=\sum_{j=1}^p J_j(x)\bar T_j(\bar x),}
and similarly for $\bar A(x,\bar x)$.
\item{(3)} Adding the operator $A(x,\bar x)$ to the spacetime Lagrangian is the same as adding to the worldsheet Lagrangian the term
\eqn\ppp{\delta\CL=\epsilon_+ K(z)\bar{J}_{\rm{SL}}^-~,}
and similarly for $\bar A(x,\bar x)$. These perturbations are irrelevant in spacetime but marginal on the worldsheet. More generally, one can add to the worldsheet Lagrangian a combination of all these couplings,
\eqn\qqq{\delta\CL_{\rm ws}=\lambda J_{\rm{SL}}^-\bar{J}_{\rm{SL}}^-+\epsilon_+ K\bar{J}_{\rm{SL}}^-+\epsilon_- J_{\rm{SL}}^-\bar K.}
On the worldsheet, this leads to a conformal manifold (a moduli space of conformal field theories). Moreover, the theory is solvable using standard techniques for all values of the couplings. In spacetime, this corresponds to deforming the block of the symmetric product \ddd, $\MM_{6k}^{(L)}$, by a combination of the corresponding irrelevant couplings,
\eqn\rrr{\delta\CL_{\rm st}=-\left(t T\bar T+\mu_+ J\bar T+\mu_- T\bar J\right).}
\item{(4)} Turning on only $\lambda$ in \qqq\ (and, therefore, only $t$ in \rrr) one gets the theory analyzed in \refs{\SmirnovLQW,\CavagliaODA,\GiveonNIE,\GiveonMYJ}. The spectrum obtained from the string theory analysis agrees with that found in field theory. Similarly, turning on only $\epsilon_+$ in \qqq\ (and, therefore, only $\mu_+$ in \rrr) gives the theory analyzed in \refs{\GuicaLIA,\ChakrabortyVJA,\ApoloQPQ}. Again, the field and string theory analyses agree. This provides strong support for the picture presented in section 2.
\item{(5)} In string theory, one can generalize the analysis of the spectrum to the case where all the couplings in \qqq\ are turned on. The discussion of this and the previous section predicts that this spectrum should agree with the field theory analysis for general values of the couplings \rrr.

\noindent
 In the next few sections, we will discuss some properties of the theory \qqq. In particular, we will describe the modifications of the $AdS_3$ geometry induced by these irrelevant deformations, the spectrum of excitations in the resulting geometries, and the relation between pathologies of the geometries and pathologies of the corresponding spectra. Our main goal in this analysis is to use irrelevant deformations of the sort \rrr\ to generalize the notion of holography beyond the AdS/CFT paradigm, and see what new issues arise.

One of the motivations for this work was that the theory with only $\epsilon_+$ turned on in \qqq, and correspondingly only $\mu_+$ turned on in \rrr, has a pathological spectrum (in the same sense as $T\bar T$ deformed CFT with negative coupling -- the energies of highly excited states are complex) for all values of the coupling. One way to fix this problem is to add in addition to the $J\bar T$ coupling a positive $T\bar T$ coupling, $\lambda$ in \qqq\ (or, equivalently, the corresponding $t$ in \rrr). If this coupling is large enough, we expect the theory to be well behaved. Thus, there must be a critical value $\epsilon_+(\lambda)$ for which unitarity is lost. It is interesting to understand the theory in the vicinity of this critical value, and the physical origin of the loss of unitarity.

\newsec{Marginal deformations of  the worldsheet sigma model}

In this section we start our analysis of the deformations \qqq\ in string theory on $AdS_3$. We will take the worldsheet $U(1)$ currents $K(z)$ and $\bar K(\bar z)$ to be associated with left and right-moving momenta on a circle labeled by $y$,
\eqn\curK{K=i\partial y,\;\;\;\bar{K}=i\bar{\partial}y.}
Another possible choice (discussed in appendix B) would be to take $K$ to be associated with left-moving momentum on one circle, $K=i\partial y_1$, and $\bar K$ to be associated with right-moving momentum on another, $\bar K=i\bar\partial y_2$. The main difference between the two choices is the spectrum of charges in the unperturbed theory. As we will see, the dependence of the energies on the charges once the perturbation \qqq\ has been turned on is not really different. The two choices of $K$, $\bar K$ also differ in the natural value of the contact term in the product $K(z)\bar K(\bar w)=C\delta^2(z-w)$. Different choices of $C$ correspond to  different choices of coordinates on the space of  theories, but do not change the space itself \KutasovXB.

More generally, our analysis is applicable to any worldsheet CFT with a $U(1)_L\times U(1)_R$ affine symmetry. States carry charges $(q_L, q_R)$ under this symmetry; the precise spectrum of charges depends on the particular CFT. We then deform the theory as in \qqq, \rrr, and compute the worldsheet sigma model and spectrum of the resulting theory. While the structure of the sigma model will depend on the detailed construction (\eg\ it will be different in the analysis of this section and that of appendix B), the dependence of the energies of states in the deformed theory on the couplings and charges of the undeformed ones will be the same, except for a possible reparametrization of the space of theories due to a different choice of the contact terms in the undeformed theory, mentioned in the previous paragraph. The reason for this is the universality of the deformations \qqq, \rrr; this universality was emphasized and used extensively in \refs{\AharonyBAD,\AharonyICS} for the special case of pure $T\bar T$ and $J\bar T$ deformations.

In spacetime supersymmetric theories there are two classes of currents we can consider. One is currents that give via the standard construction (\eg\ \refs{\GiveonNS,\KutasovZH,\KutasovXU,\GiveonJG,\BerensteinGJ,\GiveonKU}) $U(1)_R$ currents in the spacetime CFT. An example is the worldsheet $SU(2)$ currents in string theory on $AdS_3\times S^3\times T^4$. These currents give rise to the spacetime $SU(2)_R$ symmetry, which is part of the $\NN=4$ superconformal algebra. For such currents, adding the perturbations \qqq\ in general breaks supersymmetry, since the operator $A(x,\bar x)$ that is added to the spacetime Lagrangian is in this case a bottom component of a superfield w.r.t. the left-moving superalgebra, and similarly for $\bar A(x,\bar x)$, with left $\leftrightarrow$ right. A special case of \qqq\ for such currents was studied in \ApoloQPQ.

The second class of currents is one for which the operators $A$, $\bar A$ are top components of superfields, so that the perturbation \qqq\ does not break supersymmetry (of course, it does break conformal symmetry and, in general, Lorentz symmetry). An example is currents associated with the $T^4$ in $AdS_3\times S^3\times T^4$. We will mainly think of such currents below.

Our goal in this section is to find and discuss the exact worldsheet sigma-model action on $AdS_3\times S^1$  with a general deformation of the form \qqq. We will be interested in the case where the boundary is a cylinder with SUSY preserving boundary conditions. The corresponding bulk geometry is massless $BTZ\times S^1$.  This background is obtained by compactifying the spatial direction on the boundary of $AdS_3$ in Poincar\'e coordinates. In these coordinates, an element $g\in SL(2,\IR)$ is parametrized as
 \eqn\gggg{g=\pmatrix{1 & 0 \cr\gamma & 1 }
\pmatrix{e^{\phi} & 0 \cr 0 & e^{-\phi}}\pmatrix{1 & \bar{\gamma}\cr  0 & 1}
=\pmatrix{e^{\phi} & \bar{\gamma}e^{\phi}\cr
\gamma e^{\phi} & e^{-\phi}+\gamma\bar{\gamma}e^{\phi}}.}
The WZW action on $AdS_3\times S^1$ is given  by
\eqn\wzw{S=\frac{k}{2\pi}\int d^2 z \left(\partial\phi\bar{\partial}\phi+e^{2\phi}\partial\bar{\gamma}\bar{\partial}\gamma
+\frac{1}{k}\partial y\bar{\partial}y\right).}
This action is invariant under left and right-moving affine $SL(2,\IR)\times U(1)$ symmetries.

In \wzw,  $\phi$ is the radial direction, while $(\gamma,\bar{\gamma})$ parametrize the boundary, $\IR^{1,1}$.
The $M=J=0$ BTZ black hole is obtained by compactifying the spatial coordinate $\gamma_1$ on a circle of radius $R/\sqrt{\alpha'}$,
\eqn\ttxx{\gamma=\gamma_1+\gamma_0~,\qquad \bar\gamma=\gamma_1-\gamma_0~;\qquad \gamma_1\simeq\gamma_1+2\pi R/\sqrt{\alpha'}~,}
with periodic boundary conditions for all fields.

The deformation \qqq\ is shown in appendix C to give rise to the sigma model action
\eqn\edWZW{S(\lambda,\epsilon_+,\epsilon_-)=\frac{k}{2\pi} \int d^2 z\left(\partial\phi\bar{\partial}\phi+h\partial\bar{\gamma}\bar{\partial}\gamma+\frac{2\epsilon_+h}{\sqrt{k}} \partial y\bar{\partial}\gamma+\frac{2\epsilon_-h}{\sqrt{k}}\partial\bar{\gamma}\bar{\partial}y+\frac{f^{-1}h}{k}\partial y\bar{\partial}y \right),}
where
\eqn\harmf{\eqalign{f^{-1}=& \lambda + e^{-2\phi},\cr
 h^{-1}=& \lambda-4\epsilon_+\epsilon_-+e^{-2\phi}.}}
 There is also a (four dimensional) dilaton background,
 \eqn\dilaton{e^{2\Phi} = g_s^2 e^{-2\phi}h~.}
The deformed action \edWZW\ -- \dilaton\ is invariant under $U(1)_{L,\rm{null}}\times U(1)_{R,\rm{null}}\times U(1)_L\times U(1)_R$ affine symmetry, corresponding to translations of $\gamma$, $\bar\gamma$, and the left and right-moving coordinates on the circle, $y_L$ and $y_R$, respectively.

One can check that the background \edWZW\ satisfies the conformal invariance conditions (vanishing of the
worldsheet beta-functions) to leading order in $\alpha'$. In the fermionic string, this background preserves $(2,2)$  worldsheet supersymmetry (see \eg\ \GiveonCC), which implies the vanishing of the worldsheet beta-functions to all orders in $\alpha'$~\BarsSR. As mentioned above, in many cases one can perform a chiral GSO projection that leads to a spacetime supersymmetric theory \refs{\GiveonNS,\KutasovZH,\KutasovXU,\GiveonJG,\BerensteinGJ,\GiveonKU}.

Upon Kaluza-Klein (KK) reduction along the $y$ direction, one obtains the $2+1$ dimensional metric, dilaton $\Phi$, KK scalar field $\sigma$,
antisymmetric $B$-field and gauge fields $A$:
\eqn\rebg{\eqalign{ds^2=& k\left(d\phi^2+h d\gamma d\bar\gamma-fh\left(\epsilon_+d\gamma+\epsilon_-d\bar{\gamma}\right)^2\right),\cr
e^{2\Phi} =& g_s^2 e^{-2\phi}\sqrt{fh}~, \qquad e^{2\sigma}=h/f~,\cr
B_{\gamma\bar{\gamma}}=& \frac{kh}{2}~, \qquad A_\gamma= 2\sqrt{k}\epsilon_+f~,~\qquad A_{\bar{\gamma}}=2\sqrt{k}\epsilon_-f~,}}
where the functions $f, h$ are given in \harmf.

In the infrared region $\phi\to -\infty$ the background \rebg\ approaches massless BTZ, the undeformed geometry at $\lambda=\epsilon_\pm=0$. The string coupling approaches the constant value $g_s\sim1/\sqrt p$. As $\phi$ increases, the deviation from massless BTZ grows, as expected since the dual deformations of the spacetime CFT \rrr\ are irrelevant. In particular, the deformation \qqq\ changes the structure of the boundary at $\phi\to\infty$, and it is interesting to analyze the form of the resulting bulk geometry for various values of the coupling.

Looking back at \harmf, we see that for generic $\lambda$, $\epsilon_\pm$, as $\phi\to\infty$ the functions $f$ and $h$ approach constant values,
\eqn\harmfbound{f={1\over\lambda},\;\;\;  h={1\over \lambda-4\epsilon_+\epsilon_-}~.}
The background \rebg\ approaches in this limit a linear dilaton geometry with gauge field, $B$-field and scalar $\sigma$, that go to constant values at the boundary.

Two special cases of the geometry \rebg\ were discussed before:
\item{(1)} $\epsilon_\pm=0$: in this case $f=h$ \harmf. The resulting background \rebg\  appeared in \ForsteWP. In the context  of holography it was studied in \refs{\GiveonZM,\GiveonNIE}, where it was referred to as $\MM_3$. For positive $\lambda$ it smoothly interpolates between massless BTZ in the IR and a linear dilaton background in the UV. For negative $\lambda$ there is a singularity at finite $\phi$, beyond which the role of space and time on the boundary are interchanged. In that region the $\gamma_1$ circle \ttxx\ becomes a closed timelike curve.
\item{(2)} $\lambda=\epsilon_-=0$: in this case, $f(\phi)=h(\phi)=e^{2\phi}$,
and one obtains the simplest example of `warped $AdS_3$,'
which appeared \eg\ in the context of the AdS/cold atoms correspondence \refs{\SonYE,\BalasubramanianDM}, 
and as a toy model for the Kerr/CFT correspondence \AzeyanagiZD. 
The corresponding geometry \rebg\ appeared in \IsraelVV\ (see also \DetournayRH). It has closed timelike curves at large $\phi$, but no singularity \refs{\ChakrabortyVJA,\ApoloQPQ}.

\noindent
More generally, one can proceed as follows. Absence of closed timelike curves requires that for fixed $\phi$ and $\gamma_0$, the $\gamma_1$ circle is spacelike. This implies that the coefficient of $d\gamma_1^2$ in \rebg\ must be positive for all $\phi$,
\eqn\poscoeff{h(\phi)\left[1-f(\phi)(\epsilon_++\epsilon_-)^2\right]>0~.}
Consider first the case $\lambda<0$. Sending $\phi\to\infty$ in \poscoeff\ and using \harmfbound, we conclude that positivity of \poscoeff\ at large $\phi$ requires $\lambda>4\epsilon_+\epsilon_-$. Looking back at \harmf, we see that $h(\phi)$ is then positive for all $\phi$. Absence of CTC's in the geometry \rebg\ then requires that the term in square brackets in \poscoeff\ must be positive for all $\phi$ as well. But, the first line of \harmf\ says that this is not the case -- for negative $\lambda$, $f(\phi)$ diverges near the singularity at $e^{-2\phi}=|\lambda|$, and the inequality \poscoeff\ is violated in the vicinity of that point. Hence, we conclude that for $\lambda<0$ there are always CTC's in the bulk geometry.\foot{There is also a curvature singularity at  $e^{-2\phi}=|\lambda|$, as in (1) above. }

For $\lambda>0$, one has to discuss separately the cases $\lambda>(\epsilon_++\epsilon_-)^2$ and $0<\lambda<(\epsilon_++\epsilon_-)^2$. In the former case, one can show that $\lambda$ also satisfies $\lambda>4\epsilon_+\epsilon_-$, and \poscoeff\ is satisfied for all $\phi$. Hence, the geometry is regular and free of CTC's. On the other hand, for $0<\lambda<(\epsilon_++\epsilon_-)^2$ to satisfy \poscoeff\ at large $\phi$ one must have $\lambda<4\epsilon_+\epsilon_-$. Thus, $h(\phi)$ \harmf\ flips sign at a singularity at a finite value of $\phi$, while $f(\phi)$ remains positive for all $\phi$.

For $\epsilon_+\not=\epsilon_-$, \poscoeff\ is still not satisfied for all $\phi$ in this case, since $h(\phi)$ and the term in square brackets  flip sign at different values of $\phi$. For $\epsilon_+=\epsilon_-$ this problem is less severe (the closed curve is null rather than timelike at the value of $\phi$ where $h(\phi)$ flips sign), but the geometry is still pathological. Indeed, the determinant of the two dimensional metric on the boundary is proportional to $-fh$, so when $h$ flips sign one finds either two spatial directions, or a space with CTC's.

The above discussion can be summarized by saying that the qualitative behavior of \rebg\ is determined by the sign of the combination of couplings
\eqn\aaaa{\Psi=\lambda-(\epsilon_++\epsilon_-)^2.}
{\it For $\Psi>0$, $f$ and $h$ are positive for all $\phi$, and the resulting geometry  \rebg\ is smooth, with no CTC's.
For $\Psi<0$, the background \rebg\ has pathologies -- either CTC's or no timelike direction.}

It is instructive to look at some examples of the above discussion:
\item{(I)} $\epsilon_-=\epsilon_+\equiv\epsilon/2$: in this case, the metric \rebg\ takes the form
\eqn\eeqeone{ds^2=k(d\phi^2-hd\gamma_0^2+fd\gamma_1^2),}
with
\eqn\ffhhee{f^{-1}=\lambda+e^{-2\phi}~,\qquad h^{-1}=\lambda-\epsilon^2+e^{-2\phi}=\Psi+e^{-2\phi},}
and $\Psi=\lambda-\epsilon^2$, \aaaa. For $\lambda<0$, there is a CTC when $e^{2\phi}>{1\over|\lambda|}$. For $\lambda>\epsilon^2$ (\ie\ $\Psi>0$), the geometry is smooth and free of CTC's. In the intermediate region, $0<\lambda<\epsilon^2$, where $\Psi<0$,  there is a singularity at $e^{2\phi}=1/|\Psi|$, beyond which there is no timelike direction.

\item{(II)} $\epsilon_-=-\epsilon_+\equiv\epsilon/2$: the metric \rebg\ is now given by
\eqn\eeqetwo{ds^2=k(d\phi^2-fd\gamma_0^2+hd\gamma_1^2)~,}
with
\eqn\ffhheme{f^{-1}=\lambda+e^{-2\phi}~,\qquad h^{-1}=\lambda+\epsilon^2+e^{-2\phi}~,\qquad \Psi=\lambda~.}
For $\lambda>0$ the geometry is well behaved, as expected since $\Psi>0$. For $\lambda=\Psi<0$, it is pathological. There are two distinct cases to consider. If $|\lambda|<\epsilon^2$, the function $h$ \ffhheme\ is well behaved, but there is a singularity at $e^{-2\phi}=|\lambda|$, beyond which both $\gamma_0$ and $\gamma_1$ are spacelike, so there is no time. For $|\lambda|>\epsilon^2$, there is further a singularity at $e^{-2\phi}=|\lambda|-\epsilon^2$, and a CTC beyond it.

\item{(III)} $\epsilon_-=0,\, \epsilon_+\equiv\epsilon$ \GiveonFGR: the metric \rebg\ is
\eqn\ffe{ds^2=k(d\phi^2+fd\gamma d\bar\gamma-\epsilon^2 f^2 d\gamma^2)~,}
and
\eqn\fnew{f^{-1}=h^{-1}=\lambda+e^{-2\phi}~,\qquad \Psi=\lambda-\epsilon^2.}
Positivity of \poscoeff\ at large $\phi$ requires in this case $\lambda>\epsilon^2$ (\ie\ $\Psi>0$), and if this condition is satisfied, the geometry is smooth and free of CTC's for all $\phi$.

\noindent
We see that all three special cases are consistent with the general discussion above, and with the conclusion that the geometry is free of CTC's and has standard Lorentzian signature for all $\phi$ if and only if the combination of couplings $\Psi$ \aaaa\ is positive. In some cases the geometry has singularities as well, but in some others the curvature remains bounded everywhere.

The examples above illustrate another point that will play a role in our discussion below. Consider, for example, the special case $|\epsilon_+|=|\epsilon_-|=\epsilon/2$, discussed in examples I and II above. Near the boundary at $\phi\to\infty$, the metric \rebg\ takes in these cases the form \eeqeone, \eeqetwo, which can be written as
\eqn\boundmet{ds_b^2=kd\phi^2-{k\over\Psi}d\gamma_0^2+{k\over\Psi+\epsilon^2}d\gamma_1^2~.}
The fact that the dilaton in \rebg\ is linear in $\phi$ near the boundary implies that the entropy of the theory is linear in the energy at high energies (\eg\ \KutasovJP), $S=\beta_HE$.  Because of the normalization of the $d\gamma_0^2$ term in the metric \boundmet, we expect $\beta_H$ to depend non-trivially on $\Psi$,
\eqn\depbetah{\beta_H\sim\sqrt{\Psi}~,}
and in particular to go to zero as $\Psi\to 0$. Thus, the theory obtained in that limit should not have a Hagedorn growth. In the next section we will analyze the spectrum of string theory in the deformed background \edWZW\ and verify this prediction.

The geometry at $\Psi\to 0$ is intriguing. For instance, in the special cases \eeqeone, \ffhhee\ and \eeqetwo, \ffhheme,
as $\phi\to\infty$ it approaches $AdS_2\times S^1$ in Poincar\'e coordinates,
\eqn\qpqp{ds^2\to k\left(d\phi^2-e^{2\phi}d\gamma_0^2+{1\over\epsilon^2}d\gamma_1^2\right),}
with gauge fields, $B_{01},A_\mu$, and scalars $\Phi,\sigma$ whose explicit behavior is case dependent.
This means that in each sector with fixed charge under translations of $y$ and $\gamma_1$,
the dual theory will have the entropy of a certain extremal black object,
such as an extremal two dimensional charged black hole.

\newsec{Spectrum}

The main aim of this section is to derive the spectrum of string theory in the background \edWZW, obtained by deforming the theory on massless $BTZ\times S^1$ by a general perturbation of the form \qqq.  We will start by reviewing the spectrum of the undeformed theory, and then discuss the effect of the deformation. We will focus our attention on long string states, for reasons explained in section 2.

\subsec{Massless BTZ}

\noindent{\sl $\;\;5.1.1$. The spectrum of the worldsheet sigma model on massless BTZ}

The sigma-model Lagrangian on $AdS_3$ in Poincar\'e coordinates takes the form \wzw:
\eqn\adsl{\LL ={k\over2\pi}(\partial\phi\bar{\partial}\phi+ e^{2\phi}\partial\bar{\gamma}\bar{\partial}\gamma).}
The periodic identification \ttxx\ gives rise to the sigma model on massless BTZ.
It is convenient to rewrite it in the Wakimoto form
\refs{\WakimotoGF \BernardIY-\BershadskyIN},
\eqn\qWlag{\LL_W=\partial \phi \bar{\partial}\phi +\beta\bar{\partial}\gamma+\bar{\beta}\partial \bar{\gamma}-\exp\left(-\sqrt{\frac{2}{k}}\phi\right)\beta\bar{\beta}-\sqrt{\frac{2}{k}}\hat{R}\phi,}
where $\hat{R}$ is the curvature of the background metric.\foot{\qWlag\  is the worldsheet action for the fermionic string. In the bosonic string, $k$ is replaced by $k-2$.}
The Lagrangian \adsl\ is recovered upon integrating out the auxiliary fields $\beta$ and $\bar{\beta}$,
treating carefully the measure of the path integral, and rescaling the fields.
 The last term in \qWlag\ indicates that the dilaton is linear in $\phi$. The string coupling, $g_s$, behaves as $g_s\sim\exp\left(-\sqrt{\frac{1}{2k}}\phi\right)$, implying that for large $\phi$, $g_s\to 0$. Near the boundary (\ie\ as $\phi\to\infty$), the interaction term in \qWlag\ drops out, and we are left with a free Lagrangian,
\eqn\WLaga{{\cal{L}}= \beta\bar{\partial} \gamma + \bar{\beta}\partial\bar{\gamma}+\cal{L}_\phi~,}
where
\eqn\Lphia{{\cal{L}}_\phi=\partial \phi \bar{\partial} \phi-\sqrt{\frac{2}{k}}\hat{R}\phi~.}

Following \ParsonsSI, one can bosonize the $\beta-\gamma$ system as
\eqn\Wbosa{\eqalign{\gamma = i\phi_-, &  \ \ \ \ \bar{\gamma} = i\bar{\phi}_-, \cr
\beta=i\partial \phi_+, & \ \ \ \   \bar{\beta}= i\bar{\partial} \bar{\phi}_+,}}
where $\phi_{\pm}$ are defined in terms of the timelike and spacelike coordinates $\phi_0$ and $\phi_1$, respectively, as
\eqn\phipma{\phi_\pm=\frac{1}{\sqrt{2}}(\phi_0\pm \phi_1).}
The coordinates $\phi_0$ and $\phi_1$ are normalized as follows:
\eqn\phiyOPEa{\phi_\mu(z)\phi_\nu(w) \sim -\eta_{\mu\nu}\ln (z-w); \ \ \  \eta_{\mu\nu}={\rm diag}(-1,1); \ \ \ \ \mu,\nu=0,1,}
and similarly for $\bar\phi_0$, $\bar\phi_1$.

In terms of $\phi_\pm$, the Lagrangian of the free theory \WLaga\ is given by
\eqn\Wlagfa{{\cal{L}}=-\partial\phi_+\bar{\partial}\phi_--\partial\phi_-\bar{\partial}\phi_+ +{\cal L}_\phi~.}
Here $\phi_\pm$ are regular scalar fields, which have both left and right-moving components. In terms of the previously defined fields \Wbosa\ -- \phiyOPEa, the left (right)-moving component of $\phi_\pm$ in \Wlagfa\ is $\phi_\pm(\bar\phi_\pm)$ in \Wbosa. We will use the notation in \Wbosa\ below, to emphasize the relation of the fields to the original $(\beta,\gamma)$ variables.

To implement the periodic identification of the spatial coordinate on the boundary, \ttxx, we introduce the twist fields,
\eqn\twist{t^w=e^{iw(\phi_++\bar{\phi}_+)}, \ \ \ \ w\in Z,}
and impose locality of physical vertex operators with $t^w$. This leads to quantization of the spatial momentum on the boundary. Multiplying physical operators with $t^w$ leads to twisted sectors, which contain states with winding $w$ around the boundary circle.

As we will see below, physical long string states have $w\ge 1$. This can be thought of as follows. The massless BTZ background is obtained by starting with an LST background and adding to it $p$ fundamental strings, each of which winds once around the circle, with a certain orientation (see \eg\ \GiveonZM). Long string states in the $w$ twisted sector correspond to `ionizing' $w$ of the $p$ strings into a single string state with total winding $w$ and a non-zero radial momentum.

Positive $w$ corresponds to a string which winds around the circle with the same orientation as the $p-w$ remaining strings, that make up the massless BTZ background (we assume that $p\gg w$, so the background is not significantly impacted by the ``ionization''). Such a string is mutually BPS with the background, and therefore can have any radial momentum. Negative $w$ corresponds to a string with opposite orientation to the strings making up the vacuum. Such a string experiences a strong attractive force towards the background strings. Hence, there are no physical long string states in sectors with $w<0$.

 The above procedure gives vertex operators of the form\foot{It is enough for our purposes to consider string states that do not carry transverse excitations in $AdS_3$. It is possible to generalize the discussion to describe states that do.}
\eqn\vvvva{V_{BTZ}=e^{\sqrt{2\over k}j(\phi+\bar\phi)}V_{E_{L,R}}^w~,}
where
\eqn\veqwa{\eqalign{V_{E_{L,R}}^w  & =  e^{iw\phi_+ + iE_L\phi_-}e^{iw\bar{\phi}_+ + iE_R\bar{\phi}_-}\cr
& = e^{iw\frac{1}{\sqrt{2}}(\phi_0+\phi_1)+i\frac{(E+P)R}{2}{1\over\sqrt{2}}(\phi_0-\phi_1)}
  e^{iw\frac{1}{\sqrt{2}}(\bar{\phi}_0+\bar{\phi}_1)+i\frac{(E-P)R}{2}{1\over\sqrt{2}}(\bar{\phi}_0
 -\bar{\phi}_1)}.}}
$(E_L,E_R)$ are given in terms of $(E,P)$ by eq. \fff.  They are the eigenvalues of
\eqn\Wcura{J^-_{\rm SL}=\beta=i\partial\phi_+;\;\;\;  \bar{J}^-_{\rm SL} = \bar{\beta}=i\bar\partial\bar\phi_+,}
respectively. $R$ sets the scale of the problem;
one can think of it as the radius of compactification of the boundary spatial coordinate $\gamma_1$, \ttxx.

As mentioned in section 2, states which carry radial momentum have $j=-\frac{1}{2}+is$, with $s$ proportional to the momentum (see discussion around eq. \ggg). They belong to principal continuous representations of the underlying $SL(2,\IR)$ affine Lie algebra.\foot{In global $AdS_3$ one can construct normalizable states in the principal discrete $\DD^+$ ($\DD^-$) representations of $SL(2,\IR)$ \refs{\MaldacenaHW,\ArgurioTB}. In massless BTZ, where we work in the parabolic basis of $SL(2,\IR)$, the eigenvalues of $J^-_{\rm SL}$ can take any real positive (negative) value for the $\DD^+$ ($\DD^-$) representations, and the states are delta function normalizable \LindbladNigel; thus, all states are part of a continuum.}

The left and right-moving scaling dimensions $\Delta_{L,R}$ of $V_{BTZ}$ are
\eqn\DelDelbara{\Delta_{L,R}=-wE_{L,R}-\frac{j(j+1)}{k}~,\qquad \Delta_R-\Delta_L=wn~.}
We will next use them to calculate the spectrum of the spacetime theory.

\medskip

\noindent{\sl $\;\;5.1.2$. The spectrum of the spacetime theory}

Consider the type II superstring on massless $BTZ\times\cal N$, which corresponds to a Ramond ground state of the dual $CFT_2$.
Let
\eqn\vero{V_{\rm phys}=e^{-\varphi-\bar{\varphi}}V_{BTZ}V_\NN}
be a physical vertex operator of the theory in the $(-1,-1)$ picture. Here $V_{BTZ}$ is a vertex operator of the form \vvvva, and $V_\NN$ is a primary of the worldsheet $\NN=1$ superconformal algebra in the CFT $\NN$, with left and right-moving scaling dimensions $N_{L,R}$.

The on-shell condition reads:
\eqn\onshell{\Delta_{L,R}+N_{L,R}-{1\over 2}=0.}
Plugging \DelDelbara\ into \onshell, one finds the mass-shell conditions \eee. The states $V_{\rm phys}$  can be thought of as describing a string that winds $w$ times around the spatial circle in the BTZ geometry, and is moving with a certain momentum (proportional to $s$, \ggg) in the radial direction, in a particular state of transverse oscillation, described by the vertex operator $V_\NN$. Equations \eee, \fff\ give the energy and momentum on the boundary circle of such a string, in terms of its radial momentum and transverse excitation levels. Note that, as expected from our discussion above, states with positive energy only occur in sectors with $w>0$.

As described in section 2, the spectrum \eee\ is the same as that of the CFT $\left(\MM_{6k}^{(L)}\right)^p/S_p$, where the block $\MM_{6k}^{(L)}$ is the CFT associated with one long string in the BTZ background, and the winding $w$ labels the twisted sectors \refs{\GiveonMI,\ArgurioTB}.

\subsec{The spectrum of the deformed theory}

\noindent{\sl $\;\;5.2.1$. The spectrum of the deformed worldsheet sigma model}

From  the discussion in the previous subsection and section 4, it follows that at large $\phi$, the worldsheet sigma-model Lagrangian on massless $BTZ\times S^1$  in the presence of a general deformation of the form \qqq\ is given by \refs{\GiveonNIE,\ChakrabortyVJA ,\GiveonFGR}
\eqn\dlagallW{\LL=-\partial\phi_+\bar{\partial}\phi_--\partial\phi_-\bar{\partial}\phi_++\partial y\bar{\partial}y -\hat{\lambda}J^-_{\rm{}SL}\bar{J}^-_{\rm{SL}} -2\hat{\epsilon}_+K\bar{J}^-_{\rm{SL}}-2\hat{\epsilon}_-J^-_{\rm{SL}}\bar{K}+{\cal L}_\phi,}
where $\LL_{\phi}$ is given in equation \Lphia\ and the couplings $(\hat{\lambda},\hat{\epsilon}_+,\hat{\epsilon}_-)$ in \dlagallW\
and below are related to  $(\lambda,\epsilon_+,\epsilon_-)$ in \edWZW\ -- \rebg\ by \refs{\GiveonNIE,\ChakrabortyVJA ,\GiveonFGR}
\eqn\cupr{(\lambda,\epsilon_+^2,\epsilon_-^2)=\frac{R^2}{2\alpha'}(\hat{\lambda},\hat{\epsilon}_+^2,\hat{\epsilon}_-^2).}
We set the radius of the circle parametrized by $y$ to the self-dual value, $R_y=\sqrt{\alpha'}$, for simplicity. This corresponds to a particular choice of the charges of states in the undeformed theory, and is easy to generalize to arbitrary $R_y$.

The currents $J^-_{\rm{SL}}$ and $\bar{J}^-_{\rm{SL}}$ are given in \Wcura; the left and right-moving $U(1)$ currents $K$ and $\bar{K}$ are given by \curK. Thus, at large $\phi$ the deformed worldsheet Lagrangian takes the form
\eqn\Wlagf{{\cal{L}}=-\partial\phi_+\bar{\partial}\phi_--\partial\phi_-\bar{\partial}\phi_++\hat{\lambda}\partial \phi_+\bar{\partial}\phi_++2\hat{\epsilon}_+\partial y\bar{\partial}\phi_++2\hat{\epsilon}_-\partial\phi_+\bar{\partial}y +\partial y\bar{\partial}y +{\cal L}_\phi~.}
In the basis $(\phi_+,\phi_-,y)$, the background metric $G_{\mu\nu}$ and the antisymmetric $B$-field take the form
\eqn\GB{G=\pmatrix{
\hat{\lambda} & -1& (\hat{\epsilon}_++\hat{\epsilon}_- )\cr
-1 & 0 & 0\cr
(\hat{\epsilon}_++\hat{\epsilon}_-) & 0 & 1
}, \ \ \ \ B=\pmatrix{
0 & 0 & -(\hat{\epsilon}_+-\hat{\epsilon}_- )\cr
0 & 0 & 0 \cr
(\hat{\epsilon}_+-\hat{\epsilon}_- ) & 0 & 0
}.}
The analog of the vertex operators  \vvvva\ for the case of the deformed massless $BTZ\times S^1$ background take the form
\eqn\vvvv{V_{BTZ\times S^1}=e^{\sqrt{2\over k}j(\phi+\bar\phi)}V_{E_{L,R};q_{L,R}}^w~,}
where
\eqn\veqw{\eqalign{V_{E_{L,R};q_{L,R}}^w  = & e^{iw\phi_+ + iE_L\phi_- +iq_Ly}e^{iw\bar{\phi}_+ + iE_R\bar{\phi}_- +iq_R\bar{y}}\cr
 = & e^{iw\frac{1}{\sqrt{2}}(\phi_0+\phi_1)+i\frac{(E+P)R}{2}{1\over\sqrt{2}}(\phi_0-\phi_1)+i\frac{1}{\sqrt 2}(n_y-m_y)y} \cr
  \times   & e^{iw\frac{1}{\sqrt{2}}(\bar{\phi}_0+\bar{\phi}_1)+i\frac{(E-P)R}{2}{1\over\sqrt{2}}(\bar{\phi}_0
 -\bar{\phi}_1)+i\frac{1}{\sqrt 2}(n_y+m_y)\bar{y}},}}
with $(E_L,E_R)$ given by \fff, as before. The left and right-moving energies $E_{L,R}$ are now functions of the couplings $\hat{\lambda},\hat{\epsilon}_{\pm}$. The left and right-moving $U(1)$ charges are given by
\eqn\vardef{
q_L =\frac{1}{\sqrt{2}}(n_y-m_y), \ \   q_R = \frac{1}{\sqrt{2}}(n_y+m_y), }
where  $n_y,m_y\in Z$ are the quantized momentum and winding on the self-dual circle labeled by $y$.

For general values of $R_y$, one instead has $(n_y,m_y)\to(n_y\sqrt{\alpha'}/R_y,m_yR_y/\sqrt{\alpha'})$, here and below, in particular,
\eqn\genvardef{
q_L =\frac{1}{\sqrt 2}\left({n_y\sqrt{\alpha'}\over R_y}-{m_y R_y\over\sqrt{\alpha'}}\right), \ \
q_R =\frac{1}{\sqrt 2}\left({n_y\sqrt{\alpha'}\over R_y}+{m_y R_y\over\sqrt{\alpha'}}\right).}
Note that working at general $R_y$ is the same as adding an arbitrary multiple of $K\bar K$ \curK\ to the worldsheet Lagrangian \qqq, or equivalently adding an arbitrary multiple of $J\bar J$ to the spacetime one \rrr.

The problem of finding the spectrum of string theory in the background \Wlagf\ is similar in spirit to that encountered in the context of toroidal compactifications, where it gives rise to the Narain lattice (see \eg\ \GiveonFU\ for details). The slight novelty here is that one of the directions involved is timelike. Taking this into account \refs{\ChakrabortyVJA ,\GiveonFGR}, one can use the techniques developed in the context of Narain compactifications to derive the fundamental string spectrum in the background \GB.

The worldsheet dimensions of the vertex operators \vvvv, \veqw\ in the deformed theory are given by
\eqn\deltaaa{\eqalign{\Delta_{L,R} &=\frac{P_{L,R}P_{L,R}^t}{2}-\frac{j(j+1)}{k},}}
where
\eqn\pLpRa{\eqalign{&P_{L,R} =\left(n^t+m^t(B\mp G)\right)e^\ast,\cr
& e^\ast(e^\ast)^t =\frac{1}{2}G^{-1},}}
with the momentum and winding quantum numbers $n^t$ and $m^t$ given respectively by
\eqn\nma{\eqalign{n^t& =(n_+,n_-,n_y) =\frac{1}{\sqrt 2}\left(2w, \ ER, \ \sqrt{2}n_y\right),\cr
m^t & =(m_+,m_-,m_y) =\frac{1}{\sqrt 2}\left(P R, \ 0, \ \sqrt{2}m_y \right).}}
Plugging \pLpRa\ and \nma\ in \deltaaa,  the scaling dimensions of the vertex operators \vvvv\ take the form
\eqn\dimd{\eqalign{&\Delta_{L,R} = -wE_{L,R}-\frac{\hat{\lambda}}{2}E_LE_R+\frac{1}{2}(\hat{\epsilon}_+E_R+\hat{\epsilon}_-E_L)^2
+\hat{\epsilon}_+q_LE_R +\hat{\epsilon}_-q_RE_L+\frac{q_{L,R}^2}{2}-\frac{j(j+1)}{k},\cr
&\Delta_R-\Delta_L = \frac{1}{2}(q_R^2-q_L^2)+wn.}}
We can use these dimensions to obtain the spectrum of the spacetime theory.

\medskip

\noindent{\sl $\;\;5.2.2$. The spectrum of the deformed spacetime theory}

Consider the type II superstring on deformed massless $BTZ\times S^1\times \NN$, which corresponds to the Ramond ground state of the deformed $CFT_2$ \rrr. A large class of physical perturbative string states in this theory can be described by vertex operators of the form
\eqn\verp{V_{\rm phys}=e^{-\varphi-\bar{\varphi}}V_{BTZ\times S^1} V_{\cal N},}
where $V_{BTZ\times S^1}$ has the form \vvvv. As we've seen above, these operators have the same form as in the undeformed theory, but the left and right-moving energies $E_{L,R}$ are in general functions of the couplings $\hat{\lambda},\hat{\epsilon}_{\pm}$.
The scaling dimensions of $V_{BTZ\times S^1}$  after the deformation are given by \dimd, and the on-shell conditions for $V_{\rm phys}$ \verp\
are given by \onshell, where  $N_{L,R}$ are the left and right-moving scaling dimensions of $V_{\NN}$.

Following a similar line of argument used in constructing the spectrum of the spacetime theory dual to massless BTZ,
for generic $\hat{\lambda},\hat{\epsilon}_{\pm}$, one finds
\eqn\dimft{\eqalign{&h_w-\frac{kw}{4}=E_L+\frac{\hat{\lambda}}{2w}E_LE_R-\frac{1}{w}\left(\hat{\epsilon}_+q_LE_R +\hat{\epsilon}_-q_RE_L+\frac{1}{2}(\hat{\epsilon}_+E_R+\hat{\epsilon}_-E_L)^2\right),\cr
&\bar{h}_w-\frac{kw}{4}=E_R+\frac{\hat{\lambda}}{2w}E_LE_R-\frac{1}{w}\left(\hat{\epsilon}_+q_LE_R +\hat{\epsilon}_-q_RE_L+\frac{1}{2}(\hat{\epsilon}_+E_R+\hat{\epsilon}_-E_L)^2\right),\cr
&h_w-\bar{h}_w=E_L-E_R=n,}}
where $h_w,\bar h_w$ are computed in the undeformed theory (\ie\ they can be obtained by setting $\hat\lambda=\hat\epsilon_\pm=0$ in \dimft,
and using the mass-shell conditions of the undeformed theory).

As discussed in sections 2, 3, the spectrum \dimft\ is expected to agree with that of a symmetric product theory, with the block of the symmetric product given by a deformation of the long string theory \aaa\ by a general deformation of the form \rrr. In particular, for $w=1$ we expect to get the spectrum of the deformed block theory. The precise relation between the worldsheet and spacetime couplings can be read off from the special cases of this construction that were analyzed in \refs{\GiveonMYJ, \ChakrabortyVJA},
\eqn\tmumu{t=\pi\alpha'\lambda~,\qquad \mu_\pm=2\sqrt{2\alpha'}\epsilon_\pm~.}
The resulting spectrum of the deformed block theory takes the form:
\eqn\spect{ER= E_L+E_R=n+\frac{1}{2A}\left(-B-\sqrt{B^2-4AC}\right),}
with
\eqn\ABC{\eqalign{A=&\frac{1}{4}\left((\hat{\epsilon}_++\hat{\epsilon}_-)^2-\hat{\lambda}\right),\cr
B=&-1+\hat{\epsilon}_+q_L+\hat{\epsilon}_-q_R+n \hat{\epsilon}_- (\hat{\epsilon}_++\hat{\epsilon}_-)-\frac{\hat{\lambda}n}{2},\cr
C=&2\left(\bar{h}_1-\frac{c}{24}-\frac{q_R^2}{2}\right)+(q_R+n\hat{\epsilon}_-)^2;}}
and
\eqn\nncc{n=h_1-\bar{h}_1, \ \ c=6k.}
Some comments about this result are in order:
\item{(1)} While the results \spect\ -- \nncc\ were obtained for the block of a symmetric product CFT describing long strings on $AdS_3\times S^1$, \aaa, \ddd, they are applicable to any deformed CFT \rrr, due to the universality of these perturbations, \ie\ the fact that the energies of states in the deformed theory depend in a universal way on the energies, momenta and charges of the corresponding states in the undeformed theory, and on the couplings.  For special values of the couplings this was discussed in \refs{\AharonyBAD,\AharonyICS}. In appendix B we test this universality for generic values of the couplings.
\item{(2)} For pure $T\bar T$ and $J\bar T$ deformed theories, these results agree with the field theoretic analysis of \refs{\SmirnovLQW,\CavagliaODA} and \ChakrabortyVJA, respectively. For theories with both $T\bar T$ and $J\bar T$ turned on, the recent results of \LeFlochRUT\ seem to agree with \spect\ -- \nncc.
\item{(3)} An interesting question is what is the range of couplings in which the energies \spect\ -- \nncc\ are real. Since the constant $C$ is non-negative in the Ramond sector of a unitary CFT,  it's clear that for $A<0$ all energies are real. On the other hand, for $A>0$, states with large enough dimension for fixed values of the charges have complex energies in the deformed theory. This is a generalization of the phenomena found in the analysis of pure $T\bar T$ and $J\bar T$ deformed theories in \refs{\SmirnovLQW,\CavagliaODA,\ChakrabortyVJA} to all values of the couplings.
\item{(4)} Interestingly, the range of couplings in which the theory has complex energies is the same as that in which the geometry analyzed in section 4 had CTC's or no timelike direction. Indeed, there we found that the ``bad'' region was $\Psi<0$ \aaaa. Comparing the definition of $\Psi$ to $A$, we see that this is the same as $A>0$ in \ABC.
\item{(5)} Another point of comparison between the geometry of section 4 and the spectrum \spect\ -- \nncc\ concerns the high energy entropy of the theory in the ``good'' region of coupling space, $A<0$. The Cardy entropy of the undeformed CFT behaves at large dimension $h$ as $S\sim \sqrt{h}$, where we omitted the overall constant, that is easy to restore and will not be important below. In the deformed theory we see from \spect\ that the energy behaves at large $h$ as $E\sim \sqrt{\frac{C}{|A|}}$, where $C$ is proportional to $h$ (last line of \ABC).  Thus, the entropy of the deformed theory behaves at large energy as\foot{The numerical coefficient can be restored \eg\ by comparing to the pure $T\bar T$ case.}
$S\sim \sqrt{|A|}E$. This is in nice agreement with the analysis of section 4, \depbetah.
\item{(6)} It is interesting to analyze the behavior of the spectrum \spect\ -- \nncc\ in the limit $A\to 0^-$. States with $B>0$ decouple in this limit -- their energies go to $+\infty$. The energies of states with $B<0$ approach finite values, $\lim_{A\to 0^-} ER= n+\frac{2C}{|B|}$. Since $B$ depends only on the charges $(q_L,q_R,n)$ of the state, for fixed values of the charges the entropy grows like $\sqrt E$, but the coefficient of  $\sqrt E$ depends on the charges (through $B$). This is an intermediate behavior between Cardy and Hagedorn. Our construction in section 4 provides a bulk dual to this theory.

\newsec{Discussion}

The main goal of this paper was to extend the correspondence between single trace deformations of string theory on $AdS_3$ and irrelevant deformations of the dual $CFT_2$ found in \refs{\GiveonNIE,\GiveonMYJ,\ChakrabortyVJA,\ApoloQPQ} to a larger class of theories. On the string theory side, we studied the moduli space obtained by adding the generalized abelian Thirring deformations \qqq\ to the worldsheet sigma model on $AdS_3\times S^1$, which led to the background  \edWZW\ -- \dilaton. By calculating the spectrum of long fundamental strings in this background, given in eq. \spect\ -- \nncc, we obtained a prediction for a $CFT_2$ deformed by a general combination of couplings \rrr.\foot{We did not add a $J\bar J$ term in \rrr, but this is easy to do, since it corresponds to adding $K\bar K$ to the worldsheet Lagrangian \GiveonNS. The only effect of this deformation is to change the spectrum of charges in the undeformed theory. The dependence of the energies on the charges derived in section 5 remains the same. For example, in the special case \curK, the effect of the $K\bar K$ deformation is to change the undeformed charges from \vardef\ to \genvardef.} Prior work established that this prediction is correct for the cases of $T\bar T$ and $J\bar T$ deformed CFT; the recent work \LeFlochRUT\ appears to confirm the agreement for the general case as well.

One of our main results is a precise correspondence between the condition that the string background \edWZW\ -- \dilaton\ is well behaved, \ie\ has the correct signature and does not have closed timelike curves, and the condition that the spectrum \spect\ -- \nncc\ is well behaved, \ie\ all eigenstates of the Hamiltonian have real energies. In the former case, we found the condition $\Psi\geq 0$, \aaaa; in the latter, the identical condition, $A\leq 0$, \ABC.

In some cases, the geometry \edWZW\ -- \dilaton\ has curvature singularities as well as CTC's, however it seems that only the latter are related to the appearance of complex energies in the dual field theory. It would be interesting to see if theories with CTC's and complex energies can be made sense out of. If that is the case, our holographic correspondence may shed light on the way string theory resolves singularities and CTC's, and the resulting lessons can perhaps be used more broadly.

There are many other open problems related to the results of this paper. The connection between complex energies in the field theory and CTC's in the dual bulk geometry is reminiscent of the results of \CooperFFA, which pointed out a similar relation for  $T\bar T$ deformed QFT. It would be interesting to understand this connection better, and see if holography sheds any light on the fate of CTC's in string theory and field theory.

In \AharonyBAD, it was shown that the spectrum of $T\bar T$  deformed CFT can be obtained by imposing modular invariance of the torus partition sum, and using the fact that the spectrum has a universal structure -- the energies of states in the deformed theory depend in a universal way on the energies and momenta of states in the undeformed theory. In \AharonyICS, it was pointed out that this construction has a generalization to $J\bar T$ deformed CFT. The spectrum of that theory is determined by demanding modular covariance of the partition sum with a chemical potential for the charge that couples to the holomorphic conserved current $J$, and the fact that the deformed energies depend in a universal way on the energies and charges of their undeformed counterparts.

We have seen in section 5, \spect\ -- \nncc, that the universality of the spectrum generalizes to the case of arbitrary deformations of the form \rrr. Thus, it is interesting to extend the discussion of \refs{\AharonyBAD,\AharonyICS} to these theories. The main obstacle for achieving this is to understand the modular properties of the partition sum with chemical potentials coupling to all the charges. We leave this to future work \dattajiang.

In section 2, we presented some evidence for the conjecture that the CFT dual to string theory on $AdS_3$ with NS $H$-flux takes the symmetric product form \jjj, perhaps only at large $p$. It would be interesting to understand this better. This problem is also related to matrix theory descriptions of vacua of Little String Theory. As reviewed above, such vacua are described by linear dilaton backgrounds in string theory, and adding a large number of strings to these backgrounds gives a DLCQ description of the LST, in a sector with lightlike momentum equal  to the number of strings.

As is well known, the spectrum of $T\bar T$ deformed CFT $\MM$ is the same as that obtained by studying (non-)critical string theory on $\IR_t\times S^1\times\MM$ and computing the spectrum of excitations of a string wrapped once around the circle. The $T\bar T$ coupling is taken in this construction to be proportional to $\alpha'$. This observation is at first sight surprising, since we do not expect to be able to quantize strings away from the critical dimension in a Poincar\'e invariant way. However, it finds a natural home in our construction, since the transverse directions for the string are in this case described by the CFT $\MM_{6k}$ described in section 2, and the central charge of this CFT, $c_\MM=6k$, is in general non-critical. At the same time, the spectrum of such strings is given on the one hand by the standard string theory formula for wound strings, and on the other, via our construction, by $T\bar T$ deformed CFT.\foot{More support for this picture is presented in appendix D.}

For the special case $k=2$, $c_\MM=6k=12$ is the correct value for strings in the critical dimension (with $\MM$ describing the transverse space). This means that highly excited fundamental strings in the critical dimension can be described thermodynamically as BTZ black holes with that value of $k$. This seems to be related to the proposal of \GiveonPR\ of a black hole description of the thermodynamics of highly excited fundamental critical strings. It would be interesting to explore this relation further.

It is also interesting to extend the above correspondence between $T\bar T$ deformed CFT and wound strings to the larger class of theories described in this paper. In particular, it would be nice to interpret the couplings $\epsilon_\pm$ \edWZW\ from this point of view.

We saw in sections 4, 5 that, in general, in the region in coupling space where the theory is well behaved $(\Psi>0,A<0)$, the geometry is asymptotically linear dilaton, and the spectrum is asymptotically Hagedorn. However, in the limit $\Psi,A\to 0$, one gets theories that have a different behavior, both in the bulk and on the boundary. These theories deserve further attention. They appear to belong to the class of `Warped $CFT_2$'s' that plays a role in the attempts to describe Kerr black holes via the Kerr/CFT correspondence \GuicaMU\ (for a review, see \eg\ \CompereJK).

\bigskip\bigskip
\noindent{\bf Acknowledgements:}
We thank M. Asrat, N. Itzhaki and B. Kol for discussions. SC and DK thank the participants of the Simons center workshop ``$T\bar T$ and other solvable deformations of QFT's''  for many stimulating discussions.
The work of AG is supported in part by a center of excellence
supported by the Israel Science Foundation (grant number 2289/18).
The work of DK is supported in part by DOE grant DE-SC0009924.
DK thanks the Hebrew University, Tel Aviv University and the Weizmann Institute for hospitality during part of this work.

\appendix{A}{On the gap to the continuum}

From eqs. \eee, \ggg, in the $w=1$ sector, we see that the gap from the ground state to the continuum in string theory on massless $BTZ\times\NN$ is given by
\eqn\deltae{\Delta E_L=\Delta E_R={1\over 4k}+N_{\rm min}-{1\over 2}~.}
$N_{\rm min}$ is the minimal excitation level that survives the GSO projection in this theory. The duality described in section 2 predicts that \deltae\ should agree with the gap between the Ramond ground state and the bottom of the continuum in the building block of the symmetric product \ddd. This block is given by $\IR_\phi\times\NN$, \aaa, and the gap to the continuum in it depends on the properties of $\NN$. In this appendix we show that it indeed agrees with \deltae, in two extreme cases: the superstring on $AdS_3\times S^3\times T^4$, where the level $k$ can be taken to be arbitrarily large,  and on $AdS_3\times T^3$, where it is fixed to one.

In the superstring on $AdS_3\times S^3\times T^4$, where the levels of the $SL(2,\IR)$ and $SU(2)$ current algebras are taken to be $k$, the continuum \ggg\ begins at $j=-{1\over 2}$ and the lowest value of $N_{L,R}$ allowed by the chiral GSO projection is $N_{L,R}={1\over 2}$, corresponding to a single fermion excitation in the worldsheet NS sector, or by spin fields $S_\alpha$ constructed out of eight fermions (in light cone gauge) in the Ramond sector. Hence, in this case eq. \deltae\ gives rise to a gap $\Delta E_{L,R}={1\over 4k}$.

On the other hand, in the spacetime CFT, which as explained in section 2 is in this case $\MM_{6k}^{(L)}=\IR_\phi\times SU(2)_k\times T^4$, the Ramond ground state corresponds to scaling dimension $h_0={c(\MM_{6k})\over 24}={k\over 4}$, while the lowest dimension operator in the continuum has dimension $h_{\rm min}={1\over 8}(Q^{(L)})^2+{1\over 2}={k\over 4}+{1\over 4k}$.  The $1\over2$ in the first expression comes from the dimension of the spin field for all eight free fermions on $\IR_\phi\times SU(2)_k\times T^4$. In the second expression we used the value of $Q^{(L)}$ in \bbb. We see that $h_{\rm min}-h_0={1\over 4k}$, in agreement with the duality prediction.

In the superstring on $AdS_3\times T^3$, as mentioned above $k=1$, and the lowest value of the worldsheet excitation level is $N_{L,R}={1\over 4}$. In the worldsheet Ramond sector it is the dimension of a spin field for the four worldsheet fermions in light cone gauge.  In the NS sector the gap must be the same due to spacetime supersymmetry, that relates the two. Thus, \deltae\ gives in this case  $\Delta E_{L,R}=0$, \ie\ the gap to the continuum vanishes. This behavior is familiar \eg\ from  \refs{\GiveonZM,\GiveonMI}.

The long string CFT \aaa\ takes in this case the form $\MM_{6}^{(L)}=\IR_\phi\times T^3$. The Ramond ground state is at
$h_0={c(\MM_{6})\over 24}={1\over 4}$, and the continuum starts at $h_{\rm min}={1\over 8}(Q^{(L)})^2+{1\over 4}={1\over 4}$,
where the $1\over4$ in the first expression is due to the spin field of the four free fermions on $\IR_\phi\times T^3$, and in the second we used the fact that $Q^{(L)}=0$ when $k=1$, \bbb. Thus, in this case there is no gap between the ground state and the continuum in the Ramond sector of the spacetime theory, in agreement with the duality.

\appendix{B}{The spectrum of $tT\bar{T}+\mu_+J_1\bar{T}+\mu_-T\bar{J}_2$ deformed $CFT_2$}

In the main text we argued that the spectrum of the deformed theory \rrr\ is universal, \ie\ the deformed energies are given by a universal function of the undeformed energies, momenta and charges, up to a potential reparametrization of the space of theories. In this appendix we test this idea.

In section 5 we computed the spectrum of the deformed theory dual to string theory on $AdS_3\times S^1$. In that theory, the left and right-moving currents that enter the construction \rrr\ come from left and right-moving momenta on the $S^1$. Here, we examine the case where the left-moving current is associated with left-moving momentum on one $S^1$, and the right-moving one is associated with right-moving momentum on another.

Thus, we start with string theory on $AdS_3\times S^1\times S^1\times{\cal N}$, with the two circles labeled by $y_1$ and $y_2$, respectively, and deform it by
\eqn\cccb{\delta{\cal L}_{\rm ws}=\lambda J^-_{\rm{SL}}\bar{J}^-_{\rm{SL}}+\epsilon_+K_1\bar{J}^-_{\rm{SL}}
+\epsilon_-J^-_{\rm{SL}}\bar{K}_2~,}
where $K_1=i\partial y_1$ and $\bar{K}_2=i\bar\partial y_2$. As in section 5, we will take the radius of both circles to be the self dual radius, but the generalization to arbitrary radii (and in fact arbitrary metric and $B$-field on the two torus labeled by $(y_1,y_2)$) is straightforward.

In the basis $(\phi_+,\phi_-,y_1,y_2)$, the background metric $G_{\mu\nu}$ and antisymmetric $B$-field take at large $\phi$ the form
\eqn\gbd{G=\pmatrix{
\hat{\lambda}  & -1 & \hat{\epsilon}_+ & \hat{\epsilon}_- \cr
 -1 & 0 & 0 & 0 \cr
 \hat{\epsilon}_+ & 0 & 1 & 0 \cr
 \hat{\epsilon}_- & 0 & 0 & 1
}, \ \ \ \ B=\pmatrix{
0 & 0 & -\hat{\epsilon}_+ & +\hat{\epsilon} _- \cr
 0 & 0 & 0 & 0 \cr
 \hat{\epsilon} _+ & 0 & 0 & 0 \cr
 -\hat{\epsilon}_- & 0 & 0 & 0
}.}
The left and right-moving momenta along the $y_1$ circle, $(q_L,q_R)$, and those along the  $y_2$  circle, $(Q_L,Q_R)$, are given by
\eqn\momchd{\eqalign{& q_L =\frac{1}{\sqrt{2}}(n_1-m_1), \ \   q_R = \frac{1}{\sqrt{2}}(n_1+m_1),\cr
&  Q_L =\frac{1}{\sqrt{2}}(n_2-m_2), \ \   Q_R = \frac{1}{\sqrt{2}}(n_2+m_2),}}
where  $n_i,m_i\in Z$ are the momentum and winding along the $y_i$ circle $(i=1,2)$.

Following the analysis of section 5, the momentum and winding charges $n^t,m^t$ are given by
\eqn\notd{\eqalign{& n^t =(n_+,n_-,n_1,n_2) =\frac{1}{\sqrt 2}\left(2w, \ ER, \ \sqrt{2}n_1, \ \sqrt{2}n_2\right),\cr
& m^t  =(m_+,m_-,m_1,m_2) =\frac{1}{\sqrt 2}\left(P R, \ 0, \  \sqrt{2}m_1, \ \sqrt{2}m_2\right).}}
The dimensions of the low-lying vertex operators on deformed $AdS_3\times S^1\times S^1$ take the following form:
\eqn\vdimd{\eqalign{\Delta_{L,R} = &-wE_{L,R}-\frac{\hat{\lambda}}{2}E_LE_R+\frac{1}{2}(\hat{\epsilon}_+^2E_R^2+\hat{\epsilon}_-^2E_L^2)+
\hat{\epsilon}_+q_LE_R+\hat{\epsilon}_-Q_RE_L\cr
& +\frac{q_{L,R}^2}{2}+\frac{Q_{L,R}^2}{2}-\frac{j(j+1)}{k},\cr
\Delta_R-\Delta_L=& \frac{1}{2}(q_R^2-q_L^2)+\frac{1}{2}(Q_R^2-Q_L^2)+wn.}}
The spectrum of the deformed spacetime theory in the $Z_w$ twisted sector is given by
\eqn\dimftd{\eqalign{h_\omega-\frac{k\omega}{4}=&E_L+\frac{\hat{\lambda}}{2w}E_LE_R-\frac{1}{\omega}\left(\hat{\epsilon}_+q_LE_R+\hat{\epsilon}_-Q_RE_L+
\frac{1}{2}(\hat{\epsilon}_+^2E_R^2+\hat{\epsilon}_-^2E_L^2)\right),\cr
\bar{h}_\omega-\frac{k\omega}{4}=&E_R+\frac{\hat{\lambda}}{2w}E_LE_R-\frac{1}{\omega}\left(\hat{\epsilon}_+q_LE_R+\hat{\epsilon}_-Q_RE_L+
\frac{1}{2}(\hat{\epsilon}_+^2E_R^2+\hat{\epsilon}_-^2E_L^2)\right),\cr
h_\omega=\bar{h}_\omega=& E_L-E_R=n.}}
Comparing \dimftd\ to \dimft, we see that the dependence of the deformed energies on the undeformed quantum numbers is the same, except for a reparametrization of the space of theories. Indeed, taking $\hat{\lambda} \to \hat{\lambda}-2\hat{\epsilon}_-\hat{\epsilon}_+$ in \dimftd, reproduces the spectrum in \dimft. As explained in the text, this reparametrization can be traced back to a different choice of contact terms between the currents $K$ and $\bar K$ in the worldsheet theory in section 5 and here.

\appendix{C}{The deformed worldsheet sigma model}

In this appendix, we present the details of the derivation of the deformed sigma-model action \edWZW\ -- \dilaton.
The technique we use can be applied to any WZW model deformed by generalized abelian Thirring operators.

The WZW action on $AdS_3\times S^1$ is given in \wzw. The undeformed null $SL(2,\IR)_{L,R}$ currents and  $U(1)_{L,R}$ currents that will play a role in our discussion are given by
\eqn\cur{\eqalign{J_{\rm{SL}}^- = ik e^{2\phi} \partial \bar{\gamma}, & \ \ \bar{J}_{\rm{SL}}^- = ik e^{2\phi} \bar{\partial}\gamma, \cr
K= i\partial y,  &\ \  \bar{K}= i\bar{\partial}y.}}
Turning on an infinitesimal deformation of the form
\eqn\dWZW{\delta S=\frac{\delta \lambda}{2\pi k}\int d^2 z J^-_{\rm{SL}}\bar{J}^-_{\rm{SL}}-\frac{\delta\epsilon_+}{\pi\sqrt{k}}\int d^2 z
K\bar{J}^-_{\rm SL}-\frac{\delta \epsilon_-}{\pi\sqrt{k}}\int d^2 z J^-_{\rm{SL}}\bar{K},}
and analyzing the deformed action at the first few orders in the couplings $\delta\lambda,\delta\epsilon_{\pm}$,
 motivate us to define the following ansatz for the exact deformed sigma-model action at finite $\lambda,\epsilon_{\pm}$:\foot{In this appendix, $\lambda,\epsilon_\pm$ are related to those in \qqq\ by $k$-dependent factors, but agree precisely with those
that appear in the geometry \edWZW, and the rest of the paper.}
\eqn\ansatz{S(\lambda,\epsilon_+,\epsilon_-)=\frac{k}{2\pi}\int d^2 z \left(\partial\phi\bar{\partial}\phi+F\partial \bar{\gamma}\bar{\partial}\gamma
+G_1\partial\bar{\gamma}\bar{\partial}y
+G_2\partial y\bar{\partial}\gamma
+\frac{H}{k}\partial y\bar{\partial}y\right),}
where $F,G_{1,2}$ and $H$ are functions of the marginal couplings $\lambda,\epsilon_{\pm}$ and the radial coordinate $e^\phi$, to be determined.  The conserved null currents $J_{\rm{SL}}^-$ and $\bar{J}_{\rm{SL}}^-$ and the $U(1)_{L,R}$ currents $K$ and $\bar{K}$ at arbitrary $\lambda,\epsilon_{\pm}$ are given by
\eqn\dcur{\eqalign{ J_{\rm{SL}}^-(\lambda,\epsilon_+,\epsilon_-)=& ik\left(F\partial\bar{\gamma}+G_2\partial{y}\right),\cr
\bar{J}_{\rm{SL}}^-(\lambda,\epsilon_+,\epsilon_-) = & ik\left(F\bar{\partial}\gamma+G_1\bar{\partial}y\right),\cr
K(\lambda,\epsilon_+,\epsilon_-) =& ik\left(G_1\partial\bar{\gamma}+\frac{H}{k}\partial{y}\right),\cr
\bar{K}(\lambda,\epsilon_+,\epsilon_-) =& ik \left(G_2\bar{\partial}\gamma+\frac{H}{k}\bar{\partial}y\right).}}
The deformation \dWZW\ at finite $\lambda,\epsilon_{\pm}$ yields the following differential equations:
 \eqn\deqns{\eqalign{\frac{\partial S}{\partial\lambda}=&\frac{1}{2\pi k}\int d^2z J_{\rm{SL}}^-(\lambda,\epsilon_+,\epsilon_-)\bar{J}_{\rm{SL}}^-(\lambda,\epsilon_+,\epsilon_-), \cr
 \frac{\partial S}{\partial\epsilon_+}=& -\frac{1}{\pi\sqrt{k}}\int d^2 z K(\lambda,\epsilon_+,\epsilon_-)\bar{J}^-_{\rm SL} (\lambda,\epsilon_+,\epsilon_-), \cr
 \frac{\partial S}{\partial\epsilon_-}=&-\frac{1}{\pi\sqrt{k}}\int d^2 z J^-_{\rm{SL}}(\lambda,\epsilon_+,\epsilon_-)\bar{K}(\lambda,\epsilon_+,\epsilon_-).
 }}
The analysis in \refs{\ChakrabortyVJA,\GiveonFGR} provides the following initial conditions necessary to solve the above differential equations:
\eqn\bdcs{\eqalign{S(\lambda,\epsilon_+,0) = & \frac{k}{2\pi}\int d^2 z \left(\partial\phi\bar{\partial}\phi+f\partial\bar{\gamma}\bar{\partial}\gamma+2\frac{\epsilon_+f}{\sqrt{k}}\partial y\bar{\partial}\gamma +\frac{1}{k}\partial y\bar{\partial}y\right),\cr
S(\lambda,0,\epsilon_-) = & \frac{k}{2\pi}\int d^2 z \left(\partial\phi\bar{\partial}\phi+f\partial\bar{\gamma}\bar{\partial}\gamma+2\frac{\epsilon_-f}{\sqrt{k}}\partial\bar{\gamma}\bar{\partial}y +\frac{1}{k}\partial y\bar{\partial}y\right),
}}
where $f$ is defined in \harmf. The differential equations \deqns\  give rise to the following set of partial differential equations:
\eqn\pdes{\eqalign{& \frac{\partial F}{\partial\lambda}= -F^2  ,\ \  \ \ \ \ \ \  \frac{\partial G_1}{\partial\lambda}=-FG_1   ,\ \ \ \frac{\partial G_2}{\partial\lambda}=-FG_2   ,\ \ \frac{\partial H}{\partial\lambda}=-kG_1G_2   ,\cr
& \frac{\partial F}{\partial\epsilon_+}= 2\sqrt{k}FG_1  ,\ \  \frac{\partial G_1}{\partial\epsilon_+}=2\sqrt{k}G_1^2   ,\ \ \frac{\partial G_2}{\partial\epsilon_+}=\frac{2FH}{\sqrt{k}}  ,\ \ \frac{\partial H}{\partial\epsilon_+}=2\sqrt{k}HG_1   ,\cr
& \frac{\partial F}{\partial\epsilon_-}= 2\sqrt{k}FG_2  ,\ \  \frac{\partial G_1}{\partial\epsilon_-}=\frac{2FH}{\sqrt{k}},\ \ \ \frac{\partial G_2}{\partial\epsilon_-}=2\sqrt{k}G_2^2,\ \ \frac{\partial H}{\partial\epsilon_-}= 2\sqrt{k}HG_2, }}
and the initial conditions \bdcs\ give
\eqn\inc{\eqalign{F(\lambda,\epsilon_+,0)=f, \ \ G_1(\lambda,\epsilon_+,0)=0, \ \  G_2(\lambda,\epsilon_+,0)=\frac{2\epsilon_+f}{\sqrt{k}},    \ \  H(\lambda,\epsilon_+,0)=1,\cr
F(\lambda,0,\epsilon_-)=f, \ \ G_1(\lambda,0,\epsilon_-)=\frac{2\epsilon_-f}{\sqrt{k}}, \ \ G_2(\lambda,0,\epsilon_-)=0,    \ \  H(\lambda,0,\epsilon_-)=1.   }}
The partial differential equations \pdes\ with initial conditions \inc\ is solved by
\eqn\solfun{\eqalign{F(\lambda,\epsilon_+,\epsilon_-) = & h, \cr
G_1(\lambda,\epsilon_+,\epsilon_-) = & \frac{2\epsilon_-h}{\sqrt{k}}, \cr
G_2(\lambda,\epsilon_+,\epsilon_-) = & \frac{2\epsilon_+h}{\sqrt{k}}, \cr
H(\lambda,\epsilon_+,\epsilon_-) = & f^{-1}h, }}
where $f$ and $h$ are those in \harmf, and thus give rise to the exact deformed sigma-model action in \edWZW.

In the deformed theory, there is also a dilaton background, $\Phi$, which can be obtained \eg\ by recalling
that $\sqrt{|\det G|}e^{-2\Phi}$ is invariant under marginal current-current deformations (see \eg\ \refs{\GiveonJJ,\GiveonPH});
using this invariance, one gets the dilaton in equation \dilaton.

Note that the analysis \ansatz\ -- \solfun\ is reminiscent of the spacetime analysis in $T\bar T$ deformed CFT and its generalizations, in the sense that the infinitesimal perturbation \dWZW\ around a generic point in theory space is done by adding to the worldsheet Lagrangian bilinears in the currents that are conserved {\it at that point in theory space} \dcur, rather than bilinears in the original currents. This is the worldsheet analog of the fact that in the spacetime analysis we deform the theory at a generic point in the space of theories by adding to the Lagrangian a bilinear in the currents conserved at that point, and not a bilinear in the original currents, which are no longer conserved.

\appendix{D}{Insights from the null coset description}

String theory with the single trace $T\bar T$ deformation (\qqq\ with $\epsilon_+=\epsilon_-=0$)  was described in \GiveonMYJ\ as a coset of the worldsheet theory on
\eqn\nulluone{\IR_t\times S^1\times AdS_3\times\NN}
by a null $U(1)$, a linear combination of $J^-$ and a lightlike translation generator on $\IR_t\times S^1$. This description was found to be useful for understanding a number of things mentioned in the text of this paper. In this appendix we briefly review them.

\item{(1)} The quantum number $w$ in string theory on $AdS_3$ and its deformed version \qqq\ is naturally related in this construction to the usual winding number around the $S^1$ in \nulluone. This provides an alternative way of understanding the winding number of strings on $AdS_3$ found in \MaldacenaHW.

\item{(2)} The fact that the spectrum of $T\bar T$ deformed CFT is the same as the spectrum of transverse excitations of a string wrapped once around a circle, with the transverse space not necessarily critical, is a natural outcome of that construction (see \eg\ eqs. (3.1) -- (3.10) in \GiveonMYJ).

\item{(3)} The fact that the spectrum of fundamental string states in the theory arranges itself in a natural way into a symmetric product structure \jjj\ is a simple consequence of this feature for free critical strings ((3.11) -- (3.13) in \GiveonMYJ). As mentioned in the text, this feature plays an important role in matrix string theory \refs{\MotlTH,\DijkgraafVV}.

\item{(4)} The construction of \GiveonMYJ\ naturally describes states in a basis of eigenstates of $J^-$, $\bar J^-$ (the parabolic basis in  the language of \LindbladNigel). In this basis, states $|j,E\rangle$ in the $\DD^+$ ($\DD^-$) representations of $SL(2,\IR)$ are delta-function normalizable, and the spectrum is the positive (negative) real line (see \eg\ section 5 of \LindbladNigel).  Thinking of $\DD^+$ as describing in-states and $\DD^-$ as describing out-states, this feature of the spectrum is in agreement with the fact that in the theory on the cylinder with periodic boundary conditions on the circle, the energies of all states must be positive.

\listrefs
\end